 \documentclass[preprint,3p,times]{elsarticle}

\usepackage{amsmath,amssymb,amsfonts,amsthm}
\usepackage{bbm}

 \biboptions{sort&compress}

\usepackage{float}
\usepackage[export]{adjustbox}
\usepackage{placeins}
\usepackage{booktabs}

\usepackage{amsmath,bm}
\usepackage{pdflscape}
\usepackage{url}
\usepackage{textgreek}
\usepackage[final]{microtype}
\usepackage{multirow}
\usepackage{caption}
\usepackage{subcaption}
\usepackage{wrapfig}
\usepackage[symbol]{footmisc}

\usepackage[usenames]{color}

\usepackage{adjustbox}

\newcommand{\mb}[1]{\ensuremath{\mathbf{#1}}}

\journal{Powder Technology}

\begin{document}

\begin{frontmatter}

\title{Spreading of highly cohesive metal powders with transverse oscillation kinematics}

\author[MIT]{Reimar~Weissbach}
\author[MIT]{Garrett Adams} 
\author[SAM]{Patrick~M.~Praegla} 

\author[SAM]{Christoph~Meier\corref{cor1}}
\ead{christoph.anton.meier@tum.de}
\author[MIT]{A.~John~Hart\corref{cor1}}
\ead{ajhart@mit.edu}
\address[MIT]{Department of Mechanical Engineering, Massachusetts Institute of Technology, 77 Massachusetts Avenue, Cambridge, 02139, MA, USA}
\address[SAM]{Professorship of Simulation for Additive Manufacturing, Technical University of Munich, Freisinger Landstra{\ss}e 52, Garching b. M{\"u}nchen, Germany}
\address[TUM]{Institute for Computational Mechanics, Technical University of Munich, Boltzmannstra{\ss}e 15, Garching b. M{\"u}nchen, Germany}

\cortext[cor1]{Corresponding authors}

\begin{abstract}

Powder bed additive manufacturing processes such as laser powder bed fusion (LPBF) or binder jetting (BJ) benefit from using fine (D50 $\leq20~\mu m$) powders. However, the increasing level of cohesion with decreasing particle size makes spreading a uniform and continuous layer challenging. 
As a result, LPBF typically employs a coarser size distribution, and rotating roller mechanisms are used in BJ machines, that can create wave-like surface profiles due to roller run-out. 

In this work, a transverse oscillation kinematic for powder spreading is proposed, explored computationally, and validated experimentally. 
Simulations are performed using an integrated discrete element-finite element (DEM-FEM) framework and predict that transverse oscillation of a non-rotating roller facilitates the spreading of dense powder layers (beyond 50~\% packing fraction) with a high level of robustness to kinematic parameters.
The experimental study utilizes a custom-built mechanized powder spreading testbed and X-ray transmission imaging for the analysis of spread powder layers.
Experimental results generally validate the computational results, however, also exhibit parasitic layer cracking. For transverse oscillation frequencies above 200~Hz, powder layers of high packing fraction (between 50-60~\%) were formed, and for increased layer thicknesses, highly uniform and continuous layers were deposited.
Statistical analysis of the experimental powder layer morphology as a function of kinematic spreading parameters revealed that an increasing transverse surface velocity improves layer uniformity and reduces cracking defects. This suggests that with minor improvements to the machine design, the proposed transverse oscillation kinematic has the potential to result in thin and consistently uniform powder layers of highly cohesive powder.

\end{abstract}

\begin{keyword}
Metal Additive Manufacturing \sep Laser Powder Bed Fusion \sep Binder Jetting \sep Powder Spreading \sep Cohesive Powders \sep Computational Modeling  \sep Discrete Element Method
\end{keyword}

\end{frontmatter}

\section{Introduction}

Powder bed metal additive manufacturing (AM) technologies are making it possible to create intricate, high-performance components for various applications, including aerospace propulsion systems and medical implants~\cite{SpaceX2024, James2021, dias2024unveiling}. These techniques construct parts incrementally, layer-by-layer, by bonding or melting successive cross-sections. Thus, the initial step in the recurrent cycle of building a part is the powder deposition process, and the quality of the powder layer significantly influences both the quality of the part and the overall efficiency of the process.

Two of the most widespread metal additive manufacturing technologies are laser powder bed fusion (LPBF) and binder jetting (BJ). Both techniques typically involve a powder reservoir, a build platform, and one or more tools for distributing the powder. In a standard LPBF setup, the powder reservoir platform is elevated by a piston to deliver a specified amount of powder initially, while the build platform lowers by the designated nominal layer thickness to gather fresh powder. The spreading tool then moves the powder from the reservoir onto the build platform, establishing a new layer atop the previous one or the substrate. Once the powder is spread, a laser selectively melts particles in the powder bed~\cite{capozzi2022powder}. 
In BJ setups, a hopper usually replaces the piston-driven reservoir. The hopper administers the initial powder supply over the build platform or continuously in front of the spreading tool, which subsequently forms a layer on the platform. Then, a liquid binder is deposited via inkjet printing, connecting adjacent particles in specific areas to form the 'green part', which is later extracted from the printer and sintered to produce the finished part~\cite{li2020metal}.

In both processes, the powder layer influences (i) the final part's quality, since imperfections in the powder can manifest in the finished product~\cite{MIAO2022103029}, (ii) production speed, which is affected by the layer's thickness and the speed at which it is spread~\cite{li2020metal}, and (iii) the attainable geometric resolution, limited by the particle size and layer thickness. 
When spreading fine, cohesive powders as is beneficial for the processes, a counter-rotating roller is often chosen as spreading tool. One challenge when spreading with counter-rotation is the required very high manufacturing tolerance that is necessary in order to avoid for example runout that can result in a wave-like powder layer~\cite{PENNY2022Roller}. In our recent computational study, angular oscillation was identified as a promising means to deposit dense layers while avoiding challenges from roller runout~\cite{weissbach2024exploration}.
Seluga~\cite{seluga2001} initially investigated the superposition of counter-rotation with angular oscillation, revealing that when parameters are optimally chosen, this combination enhances packing fractions more effectively than mere counter-rotation. As per \cite{nasato2021influence}, vertical oscillation was investigated using DEM simulations of blade and roller spreading of PA~12 powder. The findings indicated that in the majority of cases, applying vertical oscillation to the spreading tool adversely affects the quality of the powder layer.

In this work, transverse oscillation is investigated as new kinematic approach to powder spreading. Building on promising computational results, the spreading strategy is experimentally validated using a custom-built powder spreading testbed.

\section{Computational Methods}
\label{sec:Methods}

The computational methods used in this work are the same as those applied in our previous publication focused on computational roller spreading~\cite{weissbach2024exploration}. 
The following description of computational methods is largely reused from \cite{weissbach2024exploration}.

\subsection{Computational Model}

Spreading of fine metal powders is simulated via an integrated discrete element method-finite element method (DEM-FEM) framework that was first described in \cite{Meier2019ModelingSimulations}, implemented in the parallel multi-physics research code BACI~\cite{Baci}. This model has also been applied in~\cite{Meier2019CriticalManufacturing, Meier2021GAMM, Penny2021,PENNY2022Blade,PENNY2022Roller}, but this is the first in-depth study of roller spreading for fine powders. Each particle is modeled as a discrete spherical element. Structural elements, such as the spreading tool or the substrate are modeled via FEM elements. Bulk powder dynamics are described with the resolution of individual particles following a Lagrangian approach. The model considers particle-to-particle and particle-to-wall (e.g., tool and substrate) interactions including normal contact, frictional contact, rolling resistance and cohesive forces.

Each discrete element (i.e., particle) has six degrees of freedom (i.e., three translational and three rotational DOFs) described by the position vector $\mb{r}_G$ and the rotation vector $\boldsymbol{\psi}$ and angular velocity $\boldsymbol{\omega}$ of the centerpoint of a particle. Based on these, the balance of momentum equations of a particle $i$ are~\cite{Meier2019ModelingSimulations}:
\begin{subequations}
\label{momentum}
\begin{align}
(m \, \ddot{\mb{r}}_G)^i = m^i \mb{g} + \sum_j (\mb{f}_{CN}^{ij}+\mb{f}_{CT}^{ij}+\mb{f}_{AN}^{ij}),\label{gusarov2007_HCE1}\\
(I_G \, \dot{\boldsymbol{\omega}})^i = \sum_j (\mb{m}_{R}^{ij}+\mb{r}_{CG}^{ij} \times \mb{f}_{CT}^{ij}),\label{gusarov2007_HCE2}
\end{align}
\end{subequations}
with the particle mass $m\!=\!4/3\pi r^3\rho$, the moment of inertia with respect to the particle centerpoint $I_G=0.4mr^2$, the particle radius $r$, the density $\rho$, and the gravitational acceleration $\mb{g}$. A bold symbol indicates a vector and a dot above a parameter indicates the derivative with respect to time of said parameter, e.g., $\dot{\boldsymbol{\omega}}^i$ is the angular acceleration vector of a particle $i$.
Each contact interaction between particles $i$ and $j$ is represented in Eq.~\eqref{momentum} through normal contact forces $\mb{f}_{CN}^{ij}$ (implemented with a spring-dashpot model), tangential contact forces due to Coulomb's friction $\mb{f}_{CT}^{ij}$ (with stick/slip frictional contact), adhesive forces $\mb{f}_{AN}^{ij}$, and torques $\mb{m}_{R}^{ij}$ due to rolling resistance. Furthermore, $\mb{r}_{CG}^{ij}:=\mb{r}_{C}^{ij}-\mb{r}_{G}^i$ is the vector pointing from the centerpoint of particle $i$ to the contact point with particle $j$. 

Our previous work~\cite{Meier2019ModelingSimulations, Meier2019CriticalManufacturing} showed that cohesive forces between particles dominate the dynamics and spreadability of micron-scale metal powders. 
In~\cite{Meier2019ModelingSimulations}, a cohesion force law according to the Derjaguin-Muller-Toporov (DMT) model~\cite{Derjaguin1975}, was proposed and the interested reader is referred to that reference for more details on the formulation. The surface energy $\gamma$ is a critical parameter due to (i) high uncertainty about the magnitude and variance between particles (by several orders of magnitude due to variations in particle surface roughness/topology and chemistry); and (ii) high influence on the powder behavior such as flowability and spreadability. For example, for effective surface energy values on the order of $\gamma=1$e-4$~\frac{J}{m^2}$ as identified for representative powders~\cite{Meier2019CriticalManufacturing} and mean particle diameters in the range of $d=2r \approx 30~\mu m$, cohesive forces are 10-fold greater than gravitational forces.

\subsection{Simulation setup}
\label{subsec:simulation setup}

\begin{figure}
\vspace{-12pt}
 \begin{center}
   \includegraphics[scale=1, keepaspectratio=true, width=\textwidth]{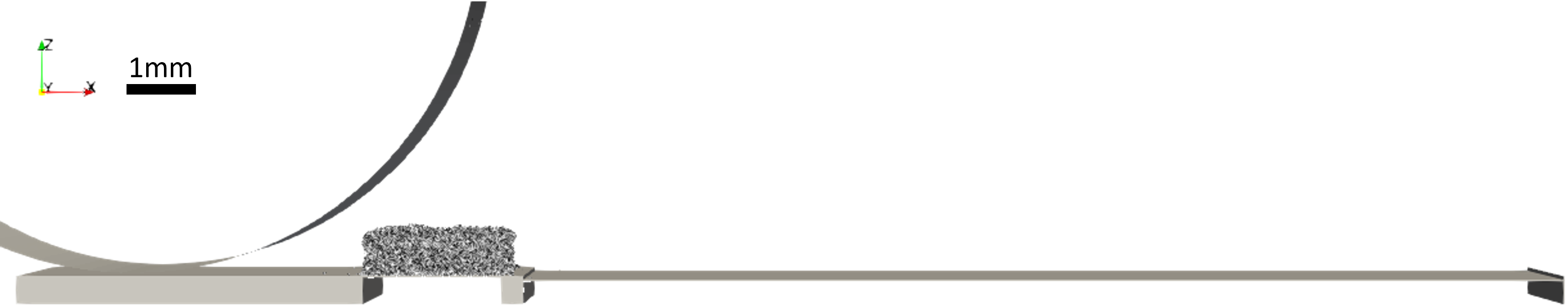}
 \end{center}
 \vspace{-16pt}
 \caption{Simulation setup with roller, fully extended powder reservoir piston with 42,000 particles and powder reservoir}
   \vspace{-12pt}
 \label{fig:simulation_setup}
\end{figure}

The following geometric setup is used for the simulation of powder spreading with a roller, with powder supplied by a reservoir as described below. The roller is modeled based on 125 linear circumferential segments of a circle with a diameter of 10~mm. For roller simulations, 42,000 particles are used. This number was chosen based on a sensitivity study and represent the threshold where a larger number did not significantly affect simulation results.
The dimensions of the powder reservoir and powder bed are similar to those used in previous work~\cite{Meier2019ModelingSimulations,Meier2019CriticalManufacturing, Meier2021GAMM, Penny2021,PENNY2022Blade,PENNY2022Roller}: the dimension perpendicular to the spreading direction is 1~mm with periodic boundary conditions, the powder bed has a length of 12~mm and quality metrics are evaluated over a 5~mm length, starting 3~mm after the beginning of the powder bed to avoid potential edge effects in the beginning or the end. The thickness of the layer is defined by the gap between the bottom edge of the spreading tool and the substrate, and is set to be $\sim$2 times the D90 particle diameter and just slightly thicker than the largest particle in the distribution (e.g., 100~$\mu m$ for a 15-45~$\mu m$ powder with a cutoff diameter of 88~$\mu m$ or $\sim$40~$\mu m$ for a 0-20~$\mu m$ powder as described below). In the beginning of the simulation, particles are randomly allocated on a grid in the powder reservoir. The particles drop down due to the gravitational force and the powder reservoir piston moves the pile of powder up and in front of the spreading tool. This situation is depicted in Figure~\ref{fig:simulation_setup}, just before the spreading tool starts to move across the powder bed.

\subsection{Quality metric: Spatially resolved packing fraction}

The primary layer quality metric analyzed in this work is the 2D packing fraction field $\Phi$ of the powder bed, quantified by its mean $\bar{\Phi}$ and standard deviation $std(\Phi)$. The computation and motivation of this metric is described in detail in \cite{Meier2019CriticalManufacturing} and will only be covered briefly here. This metric is assessed with a spatial resolution of 100x100~$\mu m$ in the length and width direction (denoted as bins), which is approximately the diameter of a typical laser spot in LPBF. For example, a low standard deviation of the packing fraction field indicates a high level of uniformity. The packing fraction is evaluated in two steps. First, each bin is further divided into smaller cubical 3D voxels with a side length of 2.5~$\mu m$ for numerical integration of the particle volume. The voxel size is chosen to minimize the error coming from the volume of particles cut by the segment walls. Finally, the packing fraction of the powder layer is evaluated as the ratio of the volume occupied by particles to the volume of the powder bed confined by the actual mean layer height. 

\textit{Remark:} For the calculation of the packing fraction, the nominal layer height is commonly employed in the literature. We do not use the nominal layer height but the actual (calculated) mean layer height. This is important since the nominal layer height usually is considerably higher than the actual mean layer height of a spread layer. The mean layer height is calculated as the mean value of the layer heights of all 100~$\mu m$ bins~\cite{Meier2019CriticalManufacturing}.

\subsection{Powder model parameters}
\label{subsec:Powder model parameters}

The particle size distribution (PSD) was fitted to experimental laser diffraction measurements in our previous work~\cite{Penny2021}, and the same values for the $10^{th}$-percentile (D10), median (D50) and $90^{th}$-percentile (D90) are fitted to a lognormal distribution in this work (D10~=~23.4~$\mu m$, D50~=~31.5~$\mu m$, D90~=~43.4~$\mu m$; cutoff diameter D$_{cutoff}$~=~88~$\mu m$). Most other powder model parameters are also kept the same as referenced in~\cite{Penny2021,PENNY2022Blade,PENNY2022Roller}, and are listed in Table~\ref{tab:DEMparameters}.

This work focuses on evaluating powders with a high level of cohesion. Cohesive forces in the model are primarily determined by the surface energy $\gamma$ of the particles. A higher value of $\gamma$ defines a more cohesive powder. In \cite{Meier2019ModelingSimulations}, we showed that the critical dimensionless group that defines the cohesiveness of a powder is the ratio of gravitational and adhesive forces, which scales according to:
\begin{align}
    \frac{F_{\gamma}}{F_g} \sim \frac{\gamma}{\rho g r^2}.
\label{eq:Gravitation to Cohesion}
\end{align}
As shown in \cite{Meier2019ModelingSimulations}, the behavior of a given powder is similar as long as the ratio stays constant. For example, a powder with a lower density, such as an aluminum alloy, and a larger size distribution might behave similar to a higher density metal, such as steel, with a smaller size distribution.
Making use of this self-similarity condition, one can model the behavior of a powder based on the known behavior of others. The present model was calibrated on the 15-45~$\mu m$ Ti-6Al-4V powder (D10~=~20.2~$\mu m$, D50~=~27.0~$\mu m$, D90~=~36.0~$\mu m$), with a surface energy value of $\gamma_0$~=~0.08~$\frac{mJ}{m^2}$~\cite{PENNY2022Roller,PENNY2022Blade}. 
In this work, we aim to model a very fine powder that has the same surface energy as the previously calibrated 15-45~$\mu m$ Ti-6Al-4V powder. 
An exemplary commercially available 0-20~$\mu m$ Ti-6Al-4V powder with D10 = 7~$\mu m$, D50 = 10~$\mu m$ and D90 = 13~$\mu m$ matches this requirement and can be modeled by retaining the size distribution of the 15-45~$\mu m$ powder, while using a modified surface energy in the model that is calculated with the transformation rule
\begin{align}
    \gamma_{fine} = \left(\frac{D50_{coarse}}{D50_{fine}}\right)^2 \times \gamma_{coarse}.
\label{eq:powder cohesion transformation}
\end{align}
Using this transformation rule, the modified surface energy takes on a value of $\gamma_{fine} = 8~\gamma_0$ or 0.64~$\frac{mJ}{m^2}$, i.e., the surface energy is increased by a factor of 8 as compared to its consistent physical value.
The same surface energy used to describe particle-particle interactions is also applied to particle-boundary interactions. 

\begin{table}
\centering
\renewcommand{\arraystretch}{0.7}
\begin{tabular}{ l l l } 
 \hline
 Parameter & Value & Unit \\ 
 \hline
 Density       &4430      &kg/m$^3$     \\ 
 Poisson's ratio     &0.342         &- \\
 Penalty parameter     &0.34         &N/m \\
 Coefficient of friction     &0.8         &- \\
 Coefficient of rolling friction     &0.07         &- \\
 Coefficient of restitution     &0.4         &- \\
 Surface energy     &0.64-1.28         &mJ/m$^2$ \\
 Hamaker constant     &$40\cdot10^{-20}$         &J \\
 Cut-off radius adhesion & $0.01$ & - \\
  \hline
 \end{tabular}
\caption{Default DEM model parameters}
\label{tab:DEMparameters}
\end{table}

Further, even though relatively cohesive metal powders have a significantly different flow behavior than coarse granular media such as dry sand, flow behavior of metal powders can be generalized more easily due to their relatively similar, high stiffness~\cite{Mandal2020}. Therefore, the findings of this study can be applied to other materials of significance to LPBF and BJ, as well as other processes requiring spreading of metal powders.

\section{Experimental Setup}
\label{sec:experimental setup}

\subsection{Mechanized Spreading Testbed}

Experimental validation of spreading strategies is performed using a modified version of a mechanized spreading testbed initially developed and built by D. Oropeza~\cite{Oropeza2021}. 
For a detailed description of the initial testbed, the interested reader is referred to \cite{Oropeza2021}.

The spreading testbed features a piston-actuated supply reservoir, a lead-screw actuated gantry that allows to move the spreading tool across the build platform with speeds up to $200\frac{mm}{s}$, and an area mimicking what would be the build platform in a LPBF machine. The build platform is a 2~mm thick, 6061 aluminum plate, only fixed on the edges to allow for X-ray transmission measurements as described later. The spreading tool is a roller with a diameter of 20~mm machined from 6061 aluminum. 
The apparatus was previously designed to accommodate spreading with a blade or a rotating roller, and in this work is modified for lateral oscillation targeting frequencies of over 1~kHz.
Thus, the rotational degree of freedom is fixed by a slot cut into one side of the roller, and inserting a 3D-printed stop that is bolted on top of the gantry. This prevents the the roller from rotating as it moves across the powder and build plate, mitigating the negative impact of roller runout observed in~\cite{PENNY2022Roller}.

\begin{figure}
 \begin{center}
   \includegraphics[trim = {0mm 0mm 0mm 0mm}, clip, scale=1, keepaspectratio=true, width=.7\textwidth]{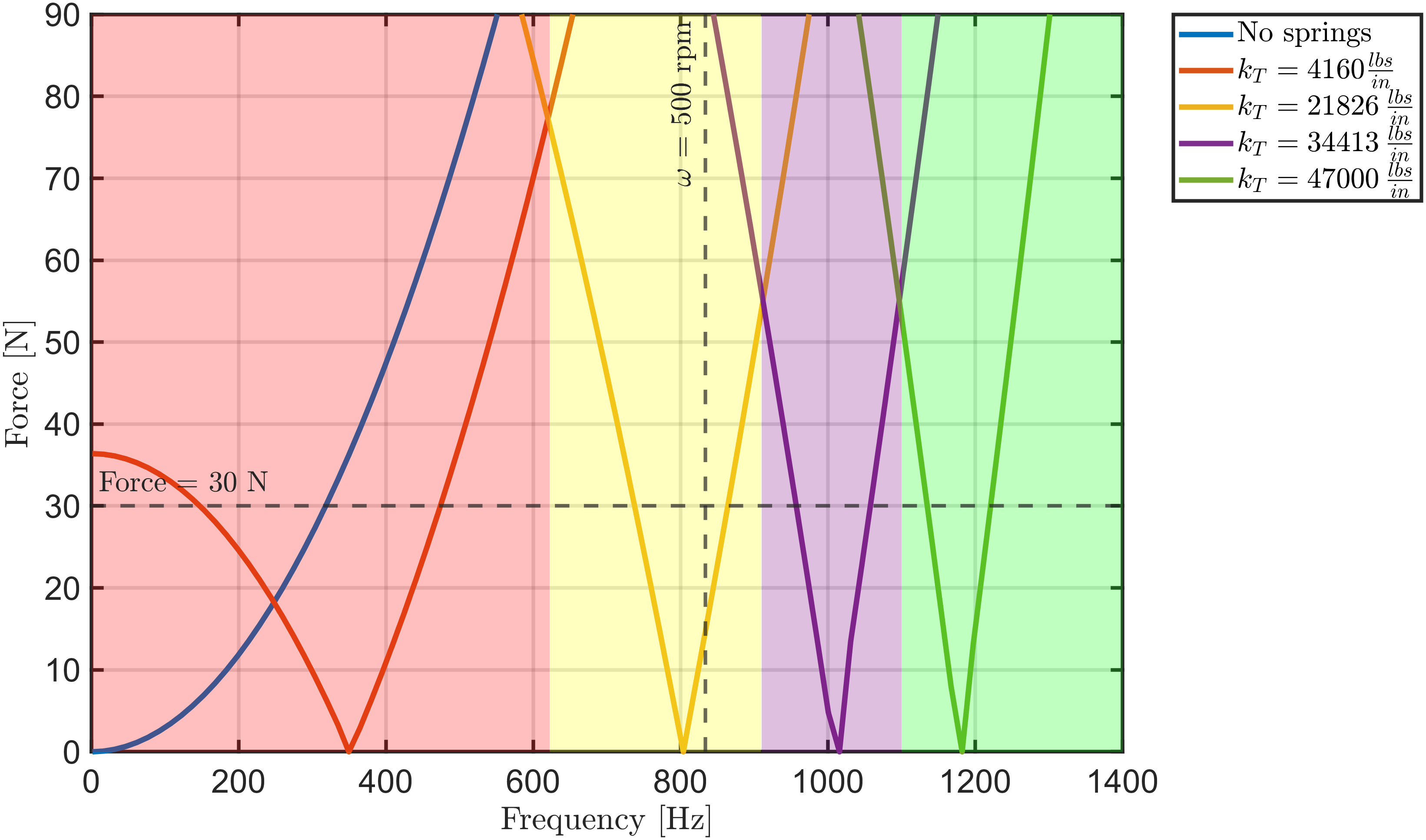}
 \end{center}
 \caption{Calculated force required to oscillate at varying frequencies and an amplitude of $50 \, \mu m$ for each combination of springs explored}
 \label{fig:force for springs}
\end{figure}

To accommodate high oscillation frequencies with a limited driving force, the roller is designed as a harmonic oscillator with pre-loaded springs on both sides. The motion can be described by a sine wave with a given frequency $\omega$ and amplitude $A$:
\begin{align}
    x = A \, \sin(\omega t).
    \label{eq: VC position}
\end{align}
With that in mind, the force required to move the roller in transverse direction is
\begin{align}
    F_{VC} = m \ddot x + b \dot x + k_T x,
    \label{eq: voice coil forces}
\end{align}
where $m$ is mass, $b$ is the damping experienced from friction, and $k_T$ is the spring constant. 
This oscillation motion is challenging because the force required to actuate the roller scales with:
\begin{align}
    F \sim m~A~(\omega - \omega_0)^2
\end{align}
when not considering damping. Thus, the required force scales quadratic with the oscillation frequency. By choosing appropriate combinations of springs and pre-loading the system, the natural frequency of the system can be designed closely to the target oscillation frequency. The employed spring combinations and frequency bands are shown in Figure~\ref{fig:force for springs}, covering a frequency band up to around 1300~Hz. 

\begin{figure}
 \begin{center}
   \includegraphics[scale=1, keepaspectratio=true, width=.4\textwidth]{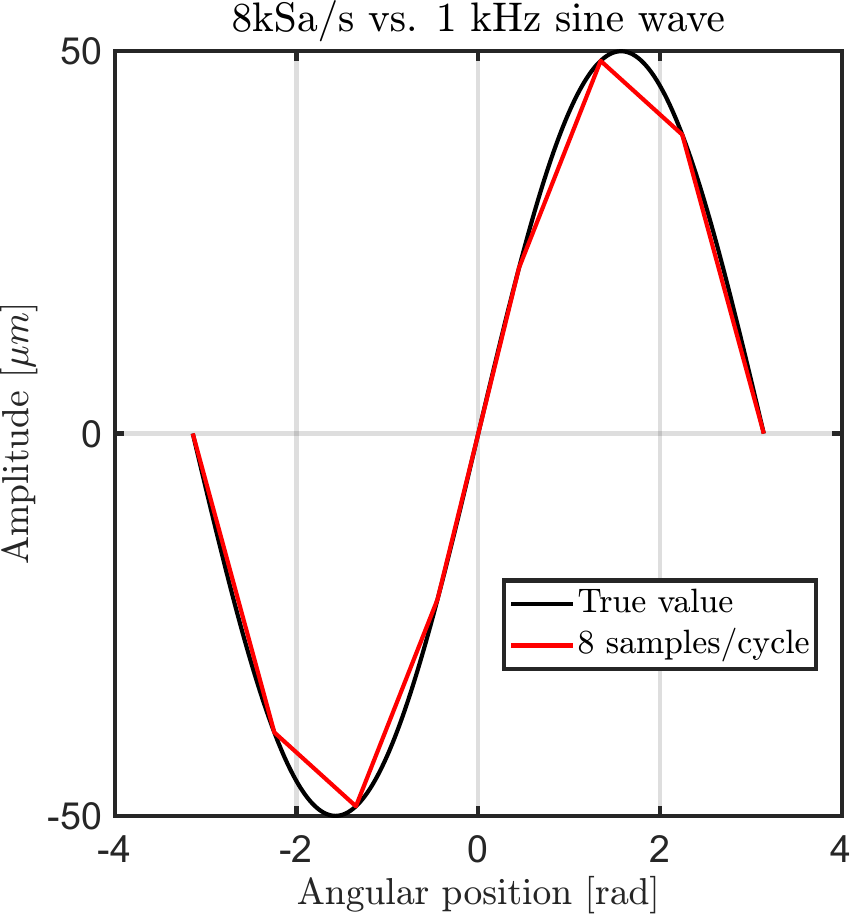}
 \end{center}
 \caption{Exemplary discretization of a sine wave with 1 kHz frequency with the maximum sample rate of the voice coil motor controller}
 \label{fig:controller samples}
\end{figure}

A voice coil motor (VCM, Moticont SDLM-051-070-01-01) was chosen as system actuator because of the high frequencies that can be reached with a reasonable force output. The VCM specifications state a maximum continuous force of 30.3~N and a maximum intermittent force of 95.7~N. The VCM has a shaft stroke of 12.7 mm, and features an internal encoder with a resolution of 1.25~$\mu m$. The VCM is controlled by a Galil DMC-30012 motor controller in a closed-loop system where programs can be written using the Galil Design Kit software. The motor controller does receive feedback from the VCM at a rate of 8~kSa/s. With that sample rate, the target frequency band up to approximately 1300~Hz can be reasonably well approximated, as shown in the exemplary discretization plot for a sine wave with 1~kHz frequency shown in Figure~\ref{fig:controller samples}.


A roller turned to a diameter of 20~mm is attached to the VCM as the spreading tool. Its derived median line (DML) straightness was measured with a Mitutoyo 293-705 digital micrometer, and was found to have a DML straightness of $\pm$ 57.15~$\mu m$. Fortunately, with a fixed roller, manufacturing defects such as tilt can be mitigated by leveling the roller.
The roughness of the roller was tested with a Keyence VK-X1000 3D microscope, and showed Ra between 2.5-3~$\mu m$ (for detailed roughness measurements, see Table~\ref{tab:rollerRoughness}).

\begin{table}
\centering
\caption{Surface roughness of the roller, as measured by the Keyence VK-X1000 3D microscope (Ra: arithmetic average of profile amplitude, Rz: maximum peak to valley height, RSm: mean width of profile elements, and R$\lambda$a: average wavelength of surface profile)}
\label{tab:rollerRoughness}
\begin{tabular}{ l l l l l l } 
\hline
 Sample Side & Ra $[\mu m]$ & Rz $[\mu m]$ & RSm $[\mu m]$ & R$\lambda$a $[\mu m]$\\ 
 \hline
 Side 1 (longitudinal)      & 2.53 & 24.83 & 1119.82 & 71.24 \\ 
 Side 1 (latitudinal) & 2.67 & 17.29 & 2535.55 & 173.29 \\
 Side 2 (longitudinal)    & 2.89 & 23.89 & 1244.45 & 79.54 \\
 Side 2 (latitudinal) & 2.78 & 17.44 & 1355.06 & 175.02 \\
\hline
\end{tabular}
\end{table}

\begin{figure}
     \centering
     \begin{subfigure}{\textwidth}
         \centering
         \includegraphics[scale=1, keepaspectratio=true, width=.8\textwidth]{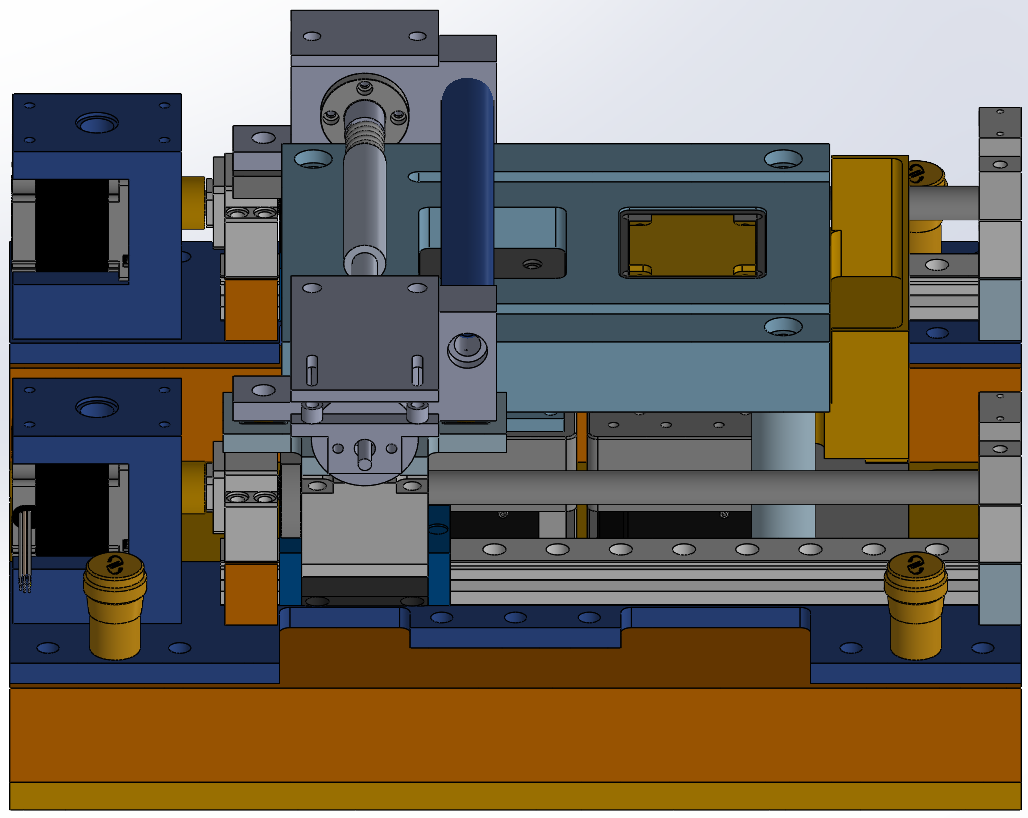}
         \caption{}
         \label{fig:Recoater iso}
     \end{subfigure}
     \vfill
     \begin{subfigure}{\textwidth}
         \centering
         \includegraphics[scale=1, keepaspectratio=true, width=.8\textwidth]{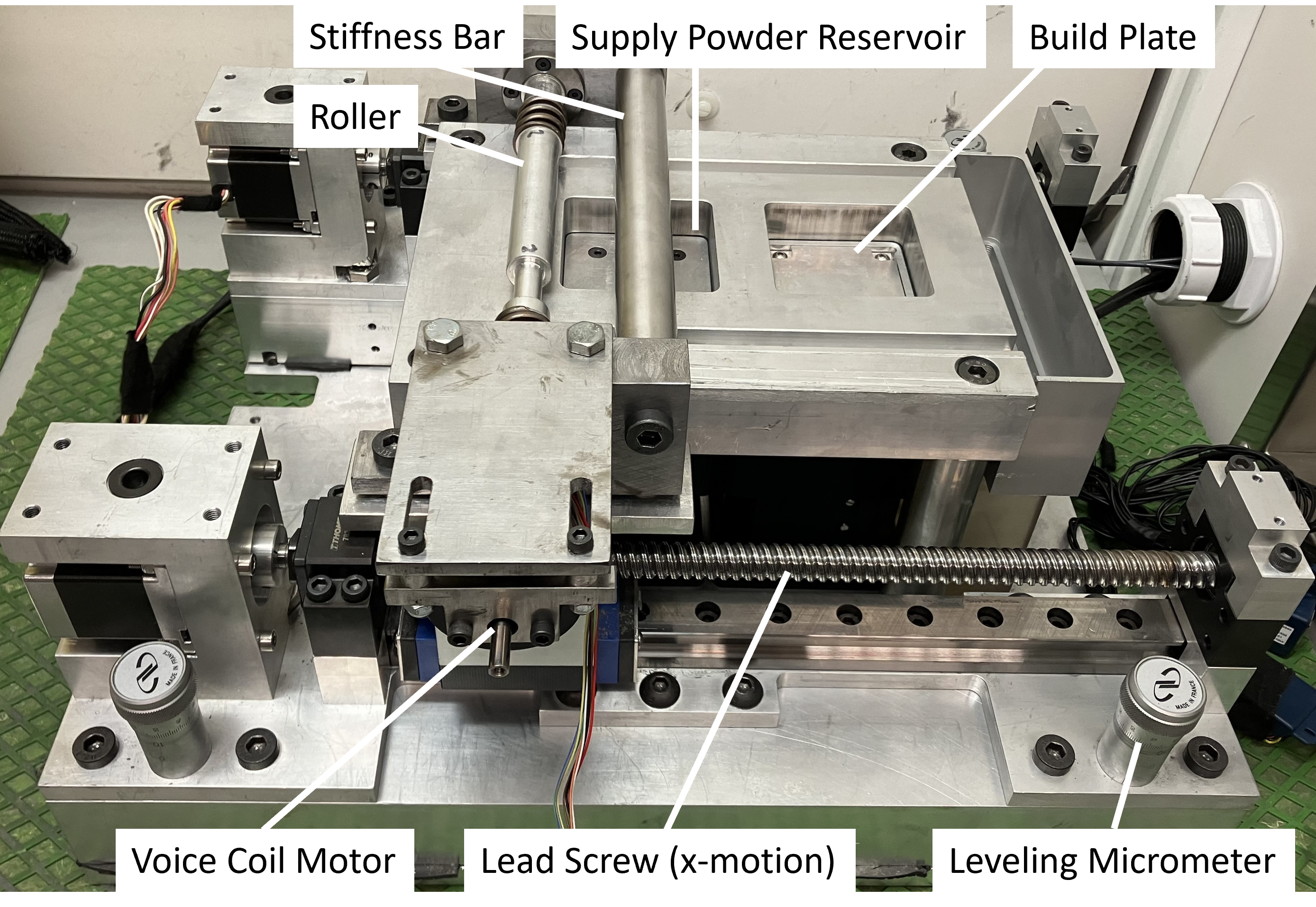}
         \caption{}
         \label{fig:Recoater side annotated}
     \end{subfigure}
     \caption{Mechanized powder spreading testbed equipped for transverse oscillation using a voice coil motor (a) CAD model including pre-loaded springs, and (b) annotated side-view of implementation}
     \label{fig:Spreading rig}
\end{figure}

\begin{figure}
 \begin{center}
   \includegraphics[width=.8\textwidth]{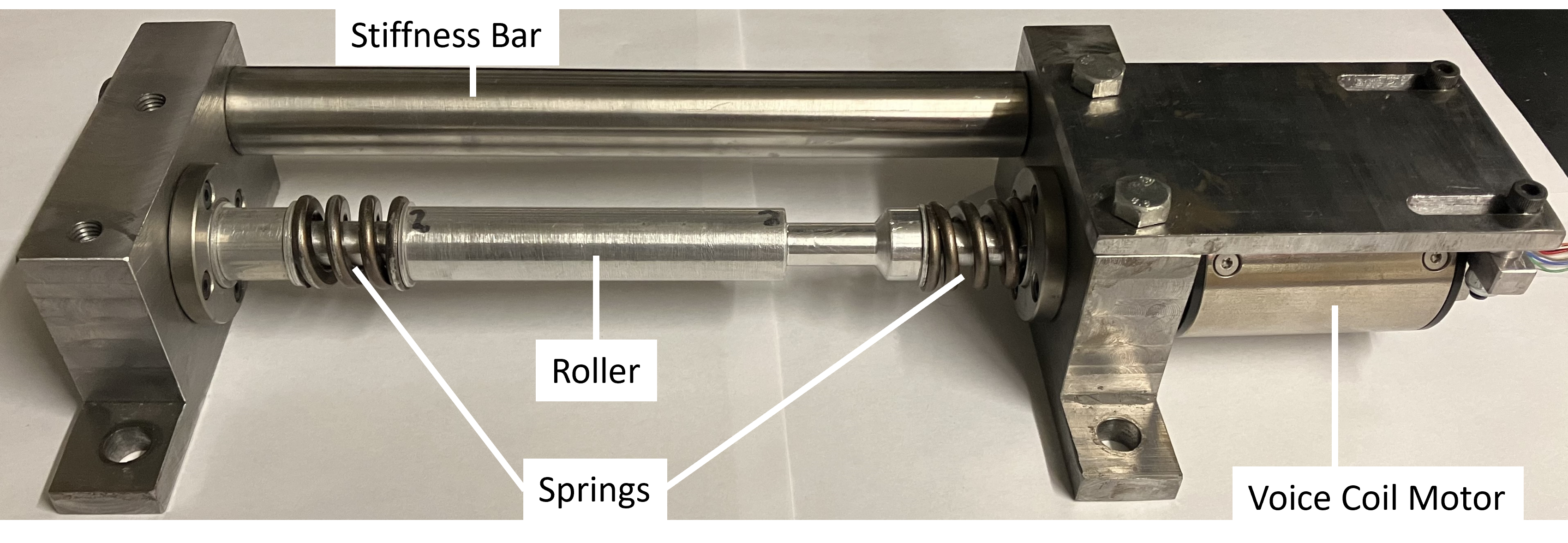}
 \end{center}
 \caption{Oscillation mechanism detached from spreading testbed}
 \label{fig:Oscillation mechanism}
\end{figure}

Figure~\ref{fig:Spreading rig} shows the CAD model (see Figure~\ref{fig:Recoater iso}), as well as the implementation of the described design as used in the study (see Figure~\ref{fig:Recoater side annotated}). Further, Figure~\ref{fig:Oscillation mechanism} shows the oscillation mechanism when detached from the spreading testbed, highlighting the roller, pre-loaded springs, and the voice coil motor.

\begin{table}
\centering
\caption{Validation data for roller movement as measured using high-speed camera videos}
\label{tab:camera_data}
\begin{tabular}{ l l l l l l p{1.2cm} p{1.3cm}} 
\toprule
 $f_{target}$ [Hz] & $\bar{f_{real}}$ [Hz] & $Err_f$ &$A_{target}$ $[\mu m]$ & $\bar{A_{real}}$ $[\mu m]$ & $Err_A$ & No. of Videos & Meas. per Video\\ 
 \hline
 200 & 204  & 2~\%  & 100 & 72  & 28~\% & 3 & 5 \\ 
 300 & 309  & 3~\%  & 100 & 55  & 45~\% & 2 & 5 \\ 
 350 & 349  & 0~\%  & 50 & 58  & 15~\% & 3 & 5 \\ 
 400 & 411  & 3~\%  & 50 & 49  & 1~\% & 3 & 5 \\ 
 450 & 455  & 1~\%  & 50 & 43  & 15~\% & 1 & 5 \\ 
 500 & 506  & 1~\%  & 50 & 34  & 33~\% & 2 & 5 \\ 
 600 & 616  & 3~\%  & 50 & 43  & 15~\% & 3 & 5 \\ 
 650 & 667  & 3~\%  & 50 & 34  & 31~\% & 3 & 5 \\ 
 700 & 711  & 2~\%  & 50 & 19  & 61~\% & 3 & 5 \\ 
 750 & 755  & 1~\%  & 50 & 15  & 71~\% & 3 & 5 \\ 
 800 & 791  & 1~\%  & 50 & 21  & 59~\% & 3 & 5 \\ 
 850 & 841  & 1~\%  & 50 & 15  & 69~\% & 3 & 5 \\ 
 900 & 897  & 0~\%  & 50 & 10  & 80~\% & 3 & 5 \\ 
 950 & 957  & 1~\%  & 50 & 14  & 72~\% & 3 & 5 \\ 
 1000 & 988  & 1~\%  & 50 & 14  & 72~\% & 3 & 5 \\ 
 1050 & 1071  & 2~\%  & 50 & 12  & 76~\% & 3 & 5 \\ 
 1100 & 1120  & 2~\%  & 50 & 10  & 79~\% & 3 & 5 \\ 
 1150 & 1173  & 2~\%  & 50 & 9  & 82~\% & 3 & 5 \\ 
 1200 & 1224  & 2~\%  & 50 & 8  & 84~\% & 2 & 5 \\ 
\hline
\end{tabular}
\end{table}

\begin{figure}
 \begin{center}
   \includegraphics[scale=1, keepaspectratio=true, width=.8\textwidth]{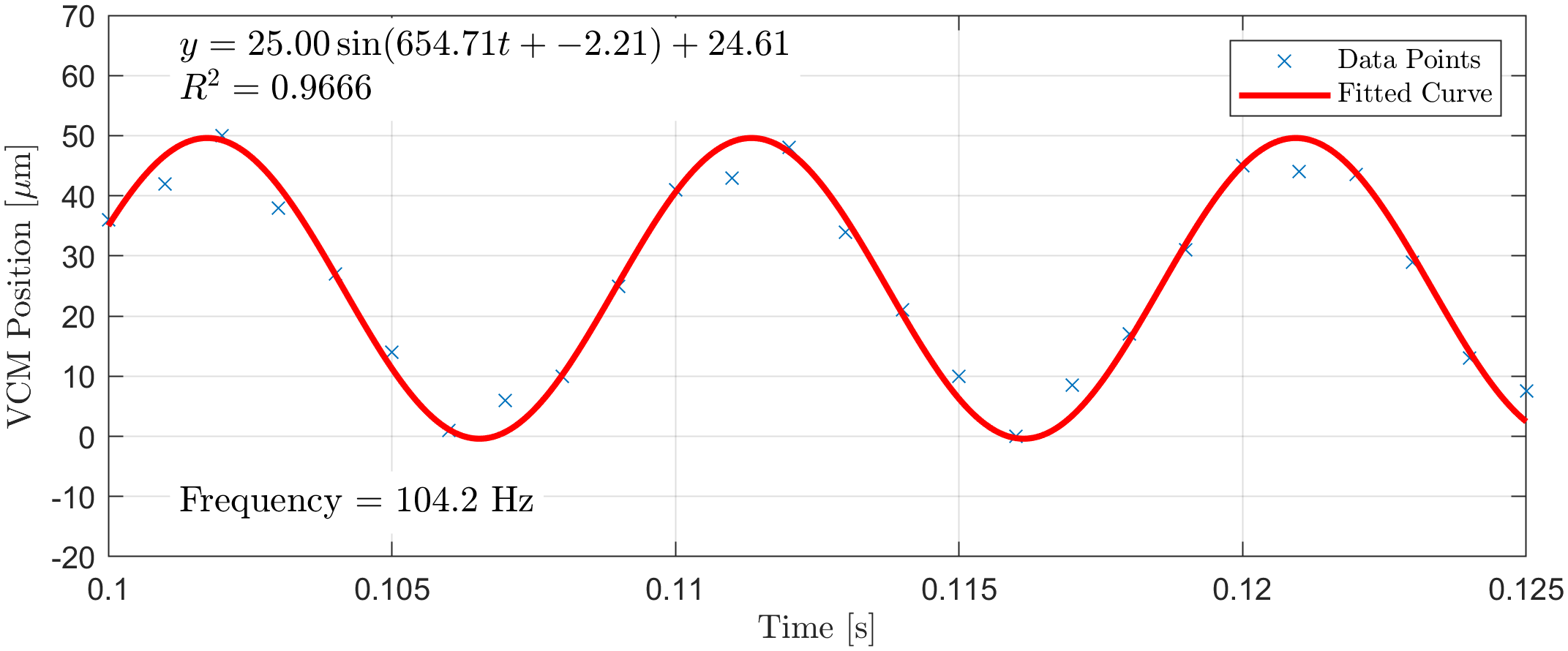}
 \end{center}
 \caption{Exemplary voice coil motor encoder output and fitted sine curve, for a target amplitude of 25$\mu m$ and frequency of 100~Hz}
 \label{fig:VCM_time_vs_position_100Hz}
\end{figure}

To validate the performance of the oscillation system, four data streams of the kinematic parameters (i.e., amplitude and frequency) were evaluated: (i) prescribed target frequency $f_{target}$ and prescribed target amplitude $A_{target}$, (ii) encoder data in controller software, (iii) encoder data captured directly via oscilloscope, and (iv) high speed camera video analysis. 
The encoder data captured via oscilloscope as well as by the controller software showed identical results, with an exemplary data stream of a sine wave with a target frequency of 100~Hz and target amplitude of $25~\mu m$ shown in Fig.~\ref{fig:VCM_time_vs_position_100Hz}.
In addition, a high-speed camera with a frame rate of up to 15~kHz was used with a resolution of ~7-22~$\mu m$ per pixel to capture the movement of the roller (specific resolution settings depending on targeted measurement). 
Using the high-speed camera, 58 videos across a range of roller frequencies up to 1200~Hz were recorded to evaluate both the frequency ($f_{real}$ with average measured frequency of a given parameter setting $\bar{f_{real}}$) and amplitude ($A_{real}$ with average measured amplitude of a given parameter setting $\bar{A_{real}}$). The amplitude and frequency of the oscillation were extracted manually from a frame-by-frame analysis of the videos (5 measurements at different points per video taken) and a summary of the measurements is shown in Table~\ref{tab:camera_data}. 
Based on the video analyses and the encoder data, the target (input) amplitude was rarely achieved by the roller, however the encoder (output) amplitude showed good agreement with the measurements from high-speed camera videos. The target (input) frequency was consistently achieved by the roller as confirmed across all three data sources. As a result, the motor encoder data stream recorded by the controller was used to assess actual amplitude and frequency of the roller during the experimental campaign.

\subsection{Powder Spreading Process}
\label{subsec:Powder Spreading Process}

The experimental procedure was as follows. The thin aluminum build plate is inserted in place of the build piston and build platform that is used outside the X-ray cabinet as described by \cite{Oropeza2021}. For each combinations of springs, the roller gantry shown in Figure~\ref{fig:Oscillation mechanism} is assembled, pre-loaded, and bolted on the lead-screw carriage as shown in Figure~\ref{fig:Recoater side annotated}. Pre-load is applied via (i) the stiffness bar, and (ii) the screws holding the roller setup to the carriage. The roller is then leveled via micrometer screws on the four corners of the lead screw assembly, and the nominal layer thickness is tested with feeler gauges and adjusted as needed. The oscillation mechanism (Figure~\ref{fig:Oscillation mechanism}) is reassembled for each set of springs, requiring frequent layer thickness verification and adjustment. 

The gap between the roller and the build platform is set to approximately 110~$\mu m$, by adjusting the height of the roller with micrometer screws until a gap gauge with 127$\mu m$ does not fit in the gap between the roller and the build plate, but a gap gauge with 102~$\mu m$ fits in with drag.

To spread a layer of powder, the roller is positioned behind the supply reservoir and an oversupply of powder (usually by a factor of approximately 3) is dispensed by raising the supply reservoir piston. The roller is actuated across the supply reservoir, pushing the powder into a pile, and forward onto the frame until it rests just on top of the closest point of the build plate to the supply reservoir. Due to setup limitations and the thin layer thickness, the build plate is raised as compared to the frame, creating a step in front of the build plate. In initial experiments it was observed that moving the roller with powder across this step induces vibrations, disturbing the powder layer.

As a result, the roller is moved forward to the edge of the build plate, without transverse oscillation activated, and then the oscillation is activated just after the roller has crossed the edge. The area that is imaged by the X-ray transmission technique is in the center of the build plate. 

The range of transverse oscillation that could be tested was limited at higher frequencies, primarily due to locking of the bearings due to misalignment of the roller ends to the linear bearings caused by the very high pre-load forces on the springs and the lack of a second stiffness (reinforcement) bar. Due to that fact, for higher frequencies, only comparatively small amplitudes of around 10~$\mu m$ were achieved consistently. The range of amplitudes that was achieved for each frequency during the experimental campaigns was in a similar range as shown in Table~\ref{tab:camera_data}.

\subsection{X-ray Microscopy of Powder Layers}
\label{subsec:chap2b X-ray method}

Powder layers were analyzed using X-ray transmission imaging in setup similar to the one employed by Penny et al.~\cite{PENNY2022Blade, PENNY2022Roller}.
This technique creates an image which represents the effective thickness which can be converted to packing density. The X-ray source is a Hamamatsu L12161-07, and the detector is a Varex 1207NDT flat-panel detector. Resolution and field of view of the final image can be defined by adjusting the distances between X-ray source, specimen stage, and detector. The resolution is set to be approximately 10~$\mu m^2$ per pixel. Each image is captured by summing over 52 frames, after subtracting dark current frames taken after every 13 images to account for noise to create a high fidelity image. The exposure integration time is 18,000~ms, and the source is set to 50~keV and 200~$\mu$A.

An image of the build plate is taken without powder, before spreading and after spreading. Each powder layer is analyzed by taking the ratio of the transmission data of the image with powder relative to the image without powder. Finally, these transmission measurements are translated into an effective depth field of the powder layer, which represents the thickness of the material if it were fully densified. The conversion parameters from radiation to effective material depth were calculated using the radiation transport model as described in \cite{Penny2021}, and using material parameters described in the following.

\subsection{Ti-6Al-4V Powder}

Experiments were carried out using Ti-6Al-4V powder supplied by AP\&C, with D10~=~7~$\mu m$, D50~=~15~$\mu m$, D90~=~24~$\mu m$. The size distribution was measured according to ASTM B822-20, with a Coulter LS Particle Size Analyzer. The tap density is $2.9\frac{g}{cm^3}$, by ASTM B527-22. 
The chemical specific chemical composition, relevant for the radiation transport model used for X-ray transmission measurements of the effective depth is as follows (all values in wt.~\%): 0.01~\% C, 0.19~\% O, 0.01~\% N, 0.005~\% H, 0.20~\% Fe, 6.36~\% Al, 4.0~\% V, $<0.2~\%$ other, balance Ti. All values are reported as measured by the supplier.

The powder is similar to the fine Ti-6Al-4V powder employed by Penny at al.~\cite{PENNY2022Roller}, where it was reported that the powder was too cohesive to flow through a hall funnel to measure the angle of repose. Further, this powder was reported to not be spreadable unless deposited with a vibrating hopper followed by a counter-rotating roller.

\section{Results}

\subsection{Computational Results}

As an alternative to the angular oscillation of the roller presented in \cite{weissbach2024exploration}, an oscillation can also be applied longitudinally to the roller, or transverse to the motion direction of the spreading tool. Therefore, the same roller parameters as in the previous studies are used, i.e., roller diameter $10$~mm, traverse velocity v~=~$25 \frac{mm}{s}$, and coefficient of friction for the roller $\mu = 0.8$. Note that this transverse oscillation approach is not limited to rollers. Instead, the roller could e.g. also be replaced by a blade with a circular edge, or a soft blade as is common in industrial applications. An equivalent displacement in transverse direction is calculated from the angular oscillation parameters via $u_y(t) = u_{y0} \sin(2\pi f t)$ with the amplitude $u_{y0}$ and frequency $f$. 

\begin{figure}
 \begin{center}
   \includegraphics[width=\textwidth]{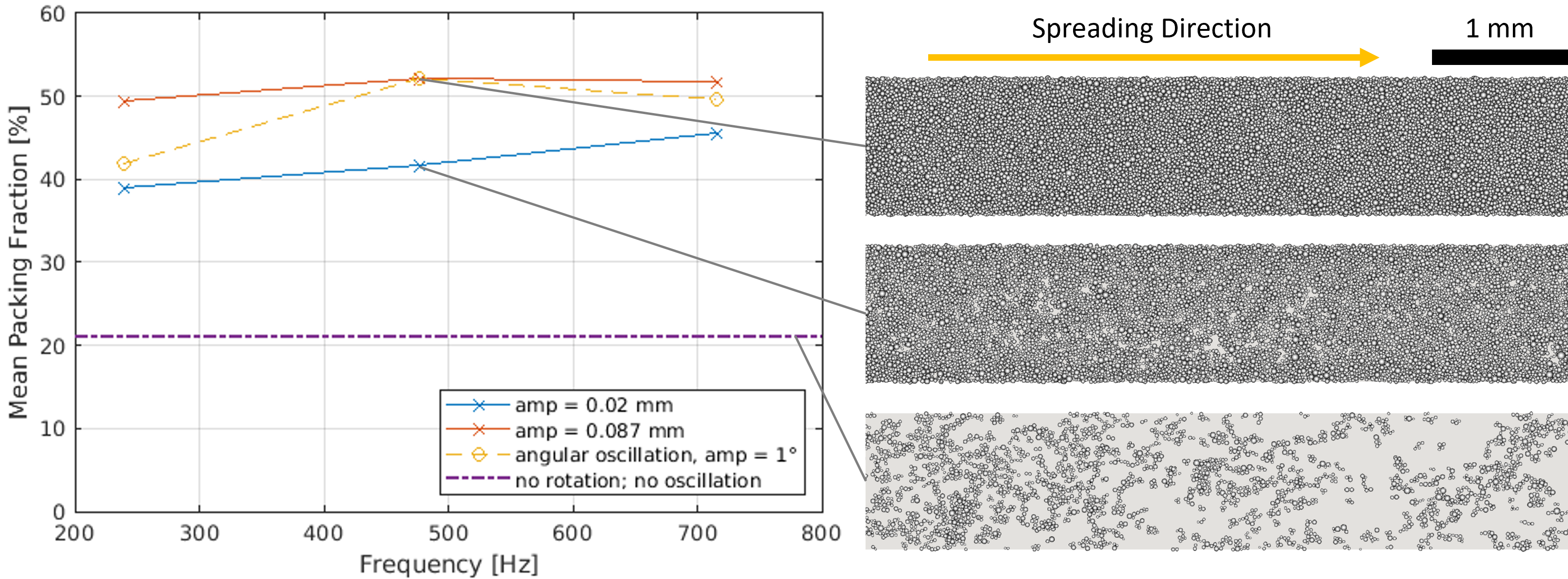}
 \end{center}
 \caption{Simulated mean packing fraction of powder layers spread with v = $25 \frac{mm}{s}$ and horizontal oscillation perpendicular to the spreading direction, for different amplitudes and frequencies and exemplary top-view of powder layers for 477~Hz transverse oscillation frequency and amplitudes of 0.087~mm, 0.02~mm, and 0~mm (from top to bottom).}
 \label{fig:roller_y_oscillation}
\end{figure}

Figure~\ref{fig:roller_y_oscillation} shows the simulated mean packing fraction of powder layers spread without any rotational or oscillation movement as baseline, compared to layers spread with angular oscillation at $1^{\circ}$ amplitude, as well as transverse oscillation at 0.02~mm and 0.087~mm amplitude. The amplitude $u_{y0} = 0.087$~mm corresponds to the same displacement as the amplitude $A = 1^{\circ}$ for the angular oscillation based on the same roller diameter, resulting in the same surface velocity profile for both oscillation patterns. 

The DEM simulations predict that oscillation results in significantly improved layer quality (40-50~\% packing fraction) compared to the baseline of no oscillation with approximately 20~\%. Within the transverse oscillation cases, the amplitude of 0.087~mm results in significantly improved powder layers with a predicted packing fraction around 50~\%, and the trajectory of layer packing fractions implies that surface velocity is the critical parameter here as well. For the same kinematic parameters, the transverse oscillation yields more robust results as compared to the angular oscillation, as the packing fraction does not vary significantly between 239~Hz and 716~Hz.

\begin{table}
\centering
\caption{Frequencies for 'equivalent maximal rotational velocity' corresponding to an amplitude of $A = 1^{\circ}$ for angular oscillation and 0.087~mm for transverse oscillation}
\label{table:oscillation parameters}
\begin{tabular}{ |p{3.7cm}|r| } 
 \hline \raggedleft
  Equivalent max rotational velocity [rpm] &Frequency [Hz]           \\ 
 \hline
 \raggedleft10    & 9.55 \\ 
 \raggedleft50    & 47.7 \\ 
 \raggedleft100   & 95.5 \\ 
 \raggedleft250   & 239 \\
 \raggedleft500   & 477 \\ 
 \raggedleft750   & 716 \\ 
 \raggedleft1000  & 955 \\
 \hline
\end{tabular}
\end{table}

\begin{figure}
     \centering
     \begin{subfigure}[b]{0.49\textwidth}
         \centering
         \includegraphics[trim = {8mm 0mm 10mm 2mm}, clip, scale=1, keepaspectratio=true, width=\textwidth]{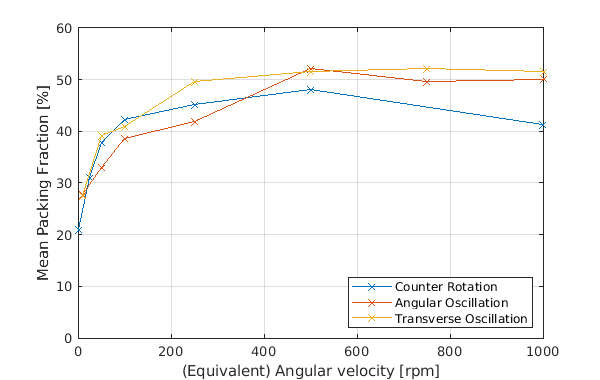}
         \caption{$\gamma = \gamma_{fine}$, $\gamma_{substrate}=\gamma$}
         \label{fig:Fig13_a}
     \end{subfigure}
     \hfill
     \begin{subfigure}[b]{0.49\textwidth}
         \centering
         \includegraphics[trim = {8mm 0mm 10mm 2mm}, clip, scale=1, keepaspectratio=true, width=\textwidth]{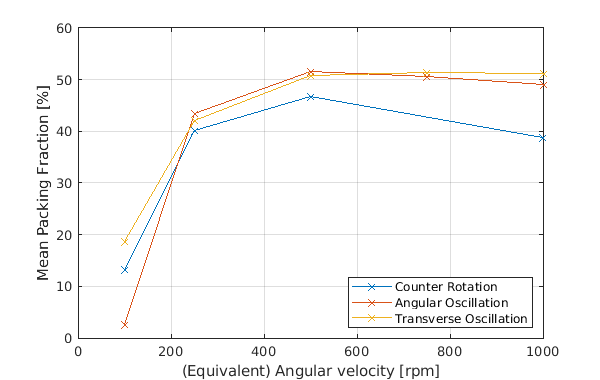}
         \caption{$\gamma = \gamma_{fine}$, $\gamma_{substrate}=0.1~\gamma$}
         \label{fig:Fig13_b_v2}
     \end{subfigure}
     \vfill
     \begin{subfigure}[b]{0.49\textwidth}
         \centering
         \includegraphics[trim = {8mm 0mm 10mm 2mm}, clip, scale=1, keepaspectratio=true, width=\textwidth]{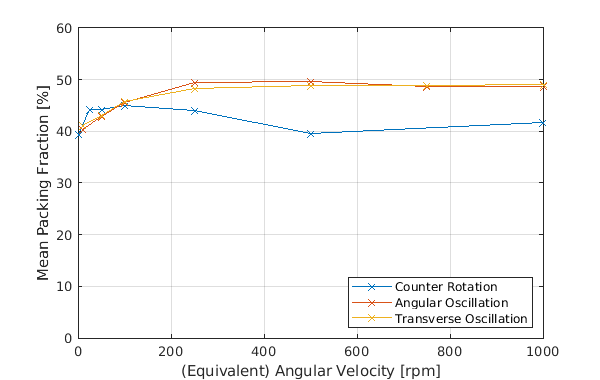}
         \caption{$\gamma = 0.5~\gamma_{fine}$, $\gamma_{substrate}=\gamma$}
         \label{fig:Paper2_comparison_low_adh_norm_padh}
     \end{subfigure}
     \hfill
     \begin{subfigure}[b]{0.49\textwidth}
         \centering
         \includegraphics[trim = {8mm 0mm 10mm 2mm}, clip, scale=1, keepaspectratio=true, width=\textwidth]{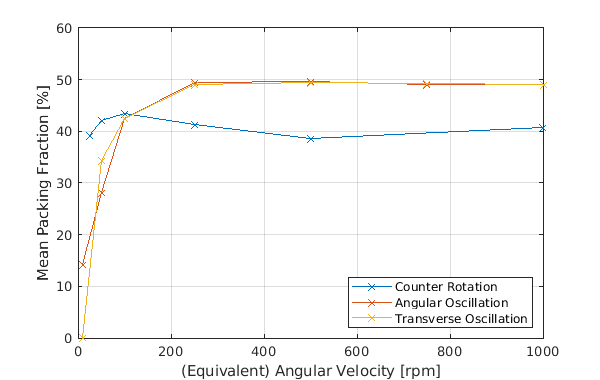}
         \caption{$\gamma = 0.5~\gamma_{fine}$, $\gamma_{substrate}=0.1~\gamma$}
         \label{fig:Paper2_comparison_low_adh_low_padh}
     \end{subfigure}
     \vfill
     \begin{subfigure}[b]{0.49\textwidth}
         \centering
         \includegraphics[trim = {8mm 0mm 10mm 2mm}, clip, scale=1, keepaspectratio=true, width=\textwidth]{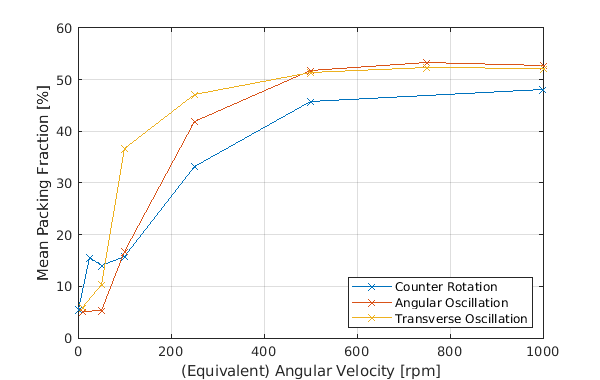}
         \caption{$\gamma = 2~\gamma_{fine}$, $\gamma_{substrate}=\gamma$}
         \label{fig:Paper2_comparison_high_adh_norm_padh}
     \end{subfigure}
     \hfill
     \begin{subfigure}[b]{0.49\textwidth}
         \centering
         \includegraphics[trim = {8mm 0mm 10mm 2mm}, clip, scale=1, keepaspectratio=true, width=\textwidth]{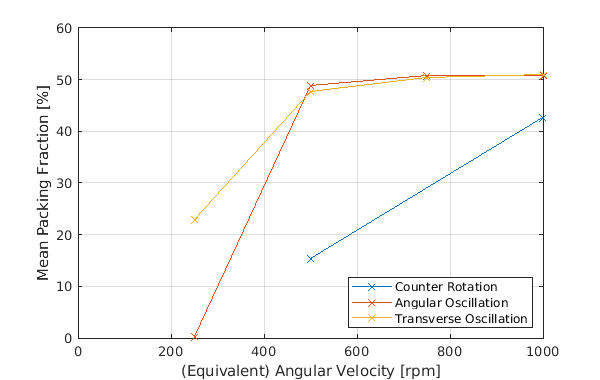}
         \caption{$\gamma = 2~\gamma_{fine}$, $\gamma_{substrate}=0.1~\gamma$}
         \label{fig:Paper2_comparison_high_adh_low_padh}
     \end{subfigure}
     \caption{Mean packing fraction of powder layers over a range of equivalent angular velocities, spread with angular counter-rotation, angular oscillation, and transverse oscillation, for varying levels of inter-particle cohesion and particle-substrate adhesion, as indicated per subfigure}
     \label{fig:spreading study transverse}
\end{figure}

To investigate robustness of the proposed spreading approaches, counter-rotation, angular oscillation, and transverse oscillation are simulated for varying levels of cohesiveness and substrate adhesiveness, with results shown in Figure~\ref{fig:spreading study transverse}.
The frequencies $f = \{9.55, 47.7, 95.5, 239, 477, 716, 955 \}$~Hz with amplitude $u_{y0} = 0.087$~mm are studied for transverse oscillation, while the equivalent displacement of $A = 1^{\circ}$ is applied for angular oscillation. The frequency-amplitude combinations are converted to their equivalent maximum rotational surface velocity for comparison to the results from the angular rotation approach. The frequency parameters correspond to what is outlined in Table~\ref{table:oscillation parameters}.

Six scenarios are analyzed in Figure~\ref{fig:spreading study transverse}: three levels of powder cohesion with $\gamma = 0.5\gamma_{fine}$ (Fig.~\ref{fig:Paper2_comparison_low_adh_norm_padh}), $\gamma = \gamma_{fine} = 0.64~\frac{mJ}{m^2}$ (Fig.~\ref{fig:Fig13_a}), and $\gamma = 2\gamma_{fine}$ (Fig.~\ref{fig:Paper2_comparison_high_adh_norm_padh}), for the case of substrate adhesion equal to interparticle cohesion, as well as the case where the substrate adhesion is reduced by a factor of 10 (Figures \ref{fig:Paper2_comparison_low_adh_low_padh}, \ref{fig:Fig13_b_v2}, and \ref{fig:Paper2_comparison_high_adh_low_padh} respectively).

As a baseline, Figure~\ref{fig:Fig13_a} shows the mean packing fraction of the layer for all three modes of spreading with $\gamma = \gamma_{fine} = \gamma_{substrate}$. All three kinematics follow a similar pattern with initially increasing powder layer quality as the (equivalent) angular velocity increases. Transverse oscillation produces the most robust results, spreading powder layers with a consistently high packing fraction across a wide range of frequencies, corresponding to angular velocities of 250~rpm to 1000~rpm. At the highest equivalent angular velocities, the layer packing fraction for both counter-rotation and angular oscillation decreases again as discussed earlier.

When the substrate surface adhesion is reduced by a factor of 10 (see Figure~\ref{fig:Fig13_b_v2}), powder layers spread with an equivalent angular velocity of 500~rpm are comparable to the baseline, but for lower velocities the layer density is substantially reduced. Similar to the baseline scenario, transverse and angular oscillation produce powder layers with a higher packing fraction than counter-rotation. Transverse oscillation furthermore has the highest level of consistency across the tested range of oscillation frequencies, a sign of robustness of the kinematic with respect to the spreading parameters as well as the properties of the substrate.

When decreasing the level of cohesion to $\gamma = 0.5\gamma_{fine}$ (Fig.~\ref{fig:Paper2_comparison_low_adh_norm_padh}), the layer quality is high across all roller surface velocities for both oscillation approaches, indicating that the maximal packing fraction in this thin single-layer simulation lies just above 50~\%. Counter-rotation produces consistent layers as well, although at a lower packing fraction compared to the oscillation. When the substrate surface energy is reduced by a factor of 10 ($\gamma_{substrate} = 0.1\gamma$), the results beyond an equivalent angular velocity of 100~rpm is comparable to the results in Fig.~\ref{fig:Paper2_comparison_low_adh_norm_padh}.
On the other end of the spectrum, for a relatively high level of cohesion ($\gamma = 2\gamma_{fine}$, see Fig.~\ref{fig:Paper2_comparison_high_adh_norm_padh}), the energy provided by oscillation or rotation of the spreading tool is predicted to be necessary to create a powder layer. At an equivalent angular velocity of 100~rpm, only transverse oscillation is able to spread a continuous and dense layer, while all three approaches manage to spread layers of high density (defined as a packing density of $>45~\%$) at 500~rpm and above. The minimum roller surface velocity for dense layers consistently increases with increasing levels of cohesion of the powder.


\subsection{Experimental Results}

Based on these promising computational results, a powder spreading testbed was modified to accommodate transverse oscillation kinematics, as described in Section~\ref{sec:experimental setup}. 
After initial validation tests in a laboratory fume hood, an experimental parameter study was conducted with the powder spreading testbed installed the X-ray apparatus, facilitating imaging of spread powder layers as described in Section~\ref{subsec:Powder Spreading Process} and Section~\ref{subsec:chap2b X-ray method}.

\begin{figure}
     \centering
     \begin{subfigure}[b]{0.49\textwidth}
         \centering
         \includegraphics[trim = {10mm 15mm 2mm 26mm}, clip, scale=1, keepaspectratio=true, width=\textwidth]{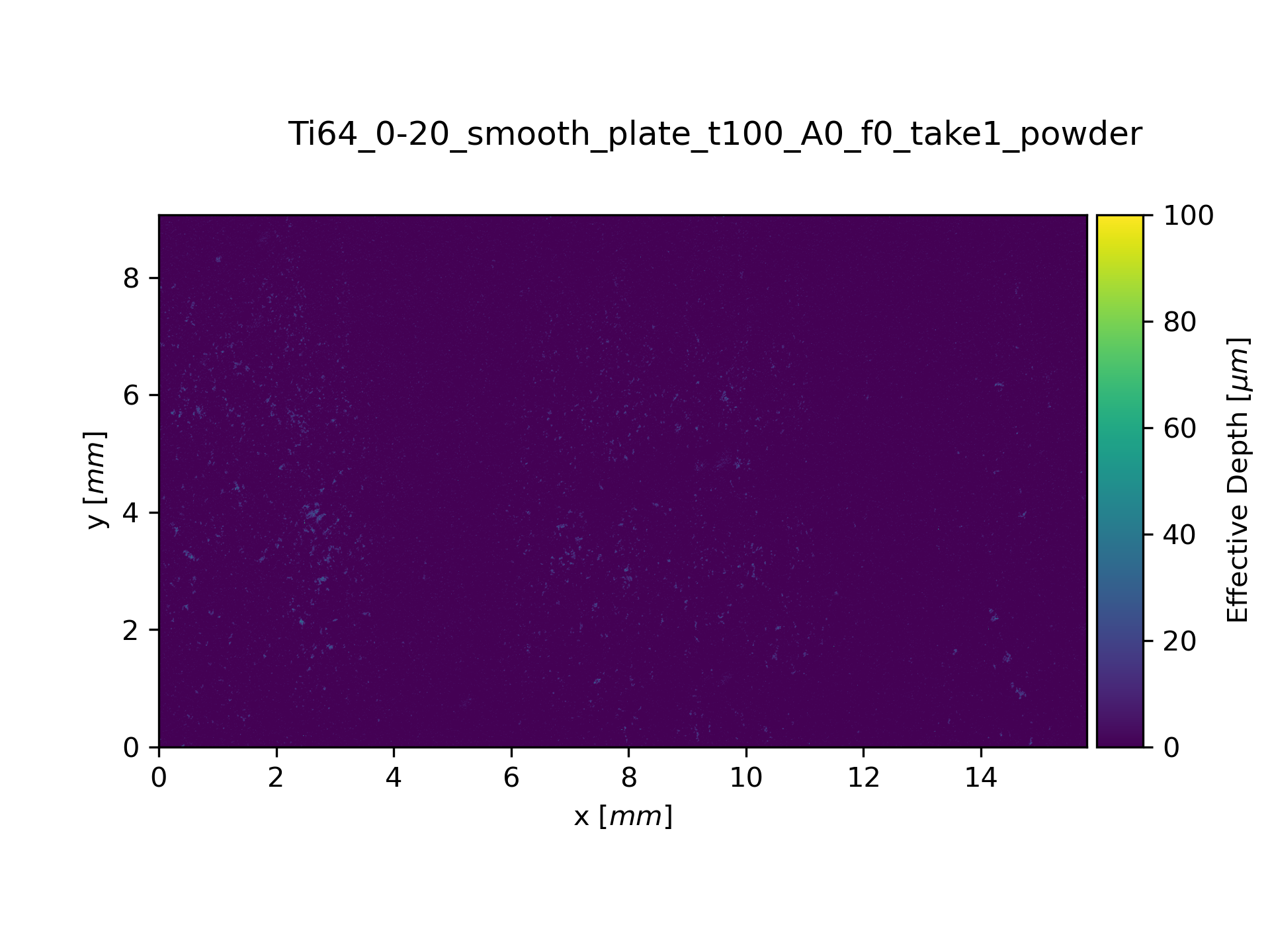}
         \caption{A=0~$\mu m$, f=0~Hz}
         \label{fig:t100_A0_f0}
     \end{subfigure}
     \hfill
     \begin{subfigure}[b]{0.49\textwidth}
         \centering
         \includegraphics[trim = {10mm 15mm 2mm 26mm}, clip, scale=1, keepaspectratio=true, width=\textwidth]{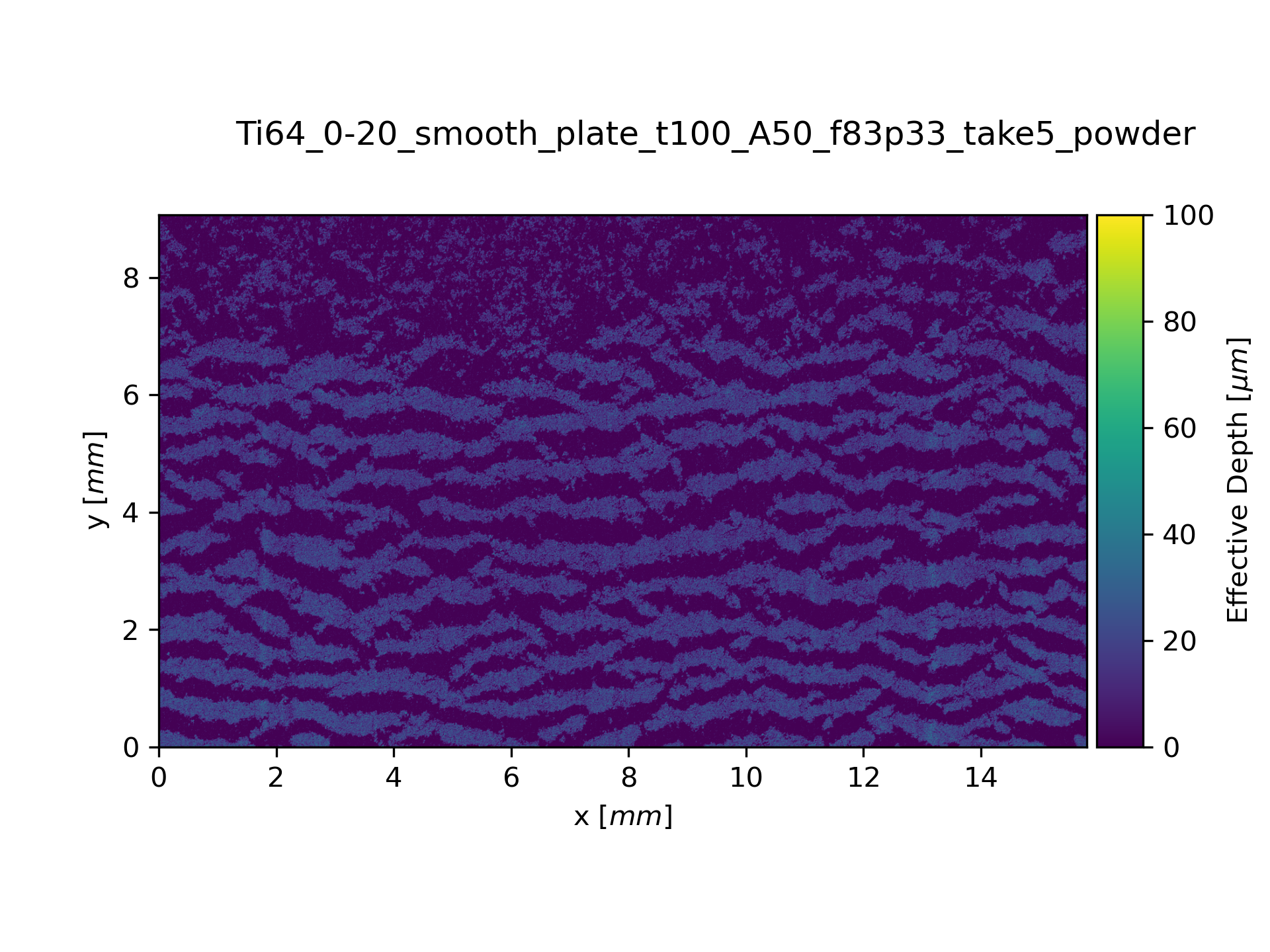}
         \caption{A=50~$\mu m$, f=83~Hz}
         \label{fig:t100_A50_f83}
     \end{subfigure}
     \vfill
     \begin{subfigure}[b]{0.49\textwidth}
         \centering
         \includegraphics[trim = {10mm 15mm 2mm 26mm}, clip, scale=1, keepaspectratio=true, width=\textwidth]{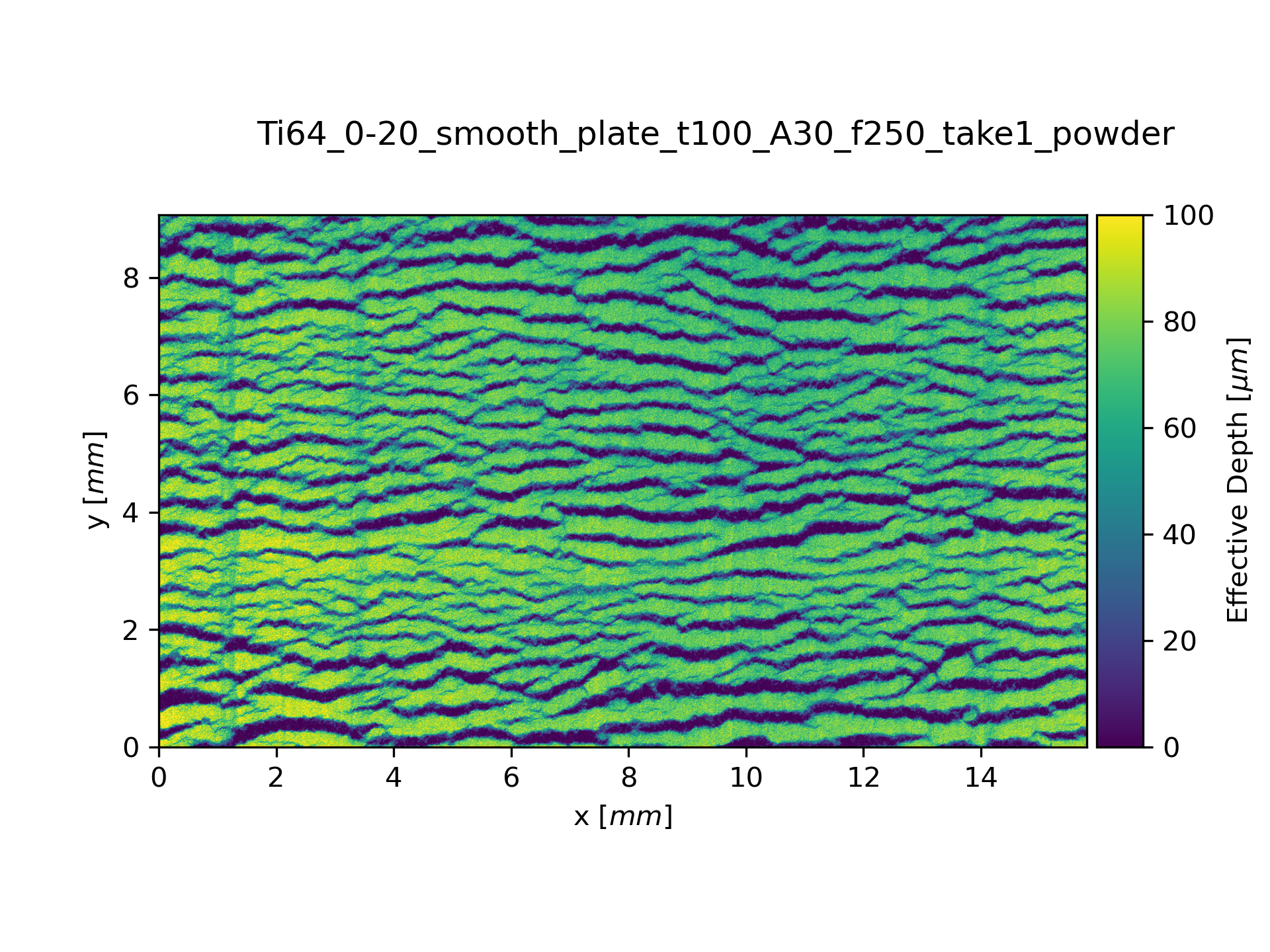}
         \caption{A=30~$\mu m$, f=250~Hz}
         \label{fig:t100_A30_f250}
     \end{subfigure}
     \hfill
     \begin{subfigure}[b]{0.49\textwidth}
         \centering
         \includegraphics[trim = {10mm 15mm 2mm 26mm}, clip, scale=1, keepaspectratio=true, width=\textwidth]{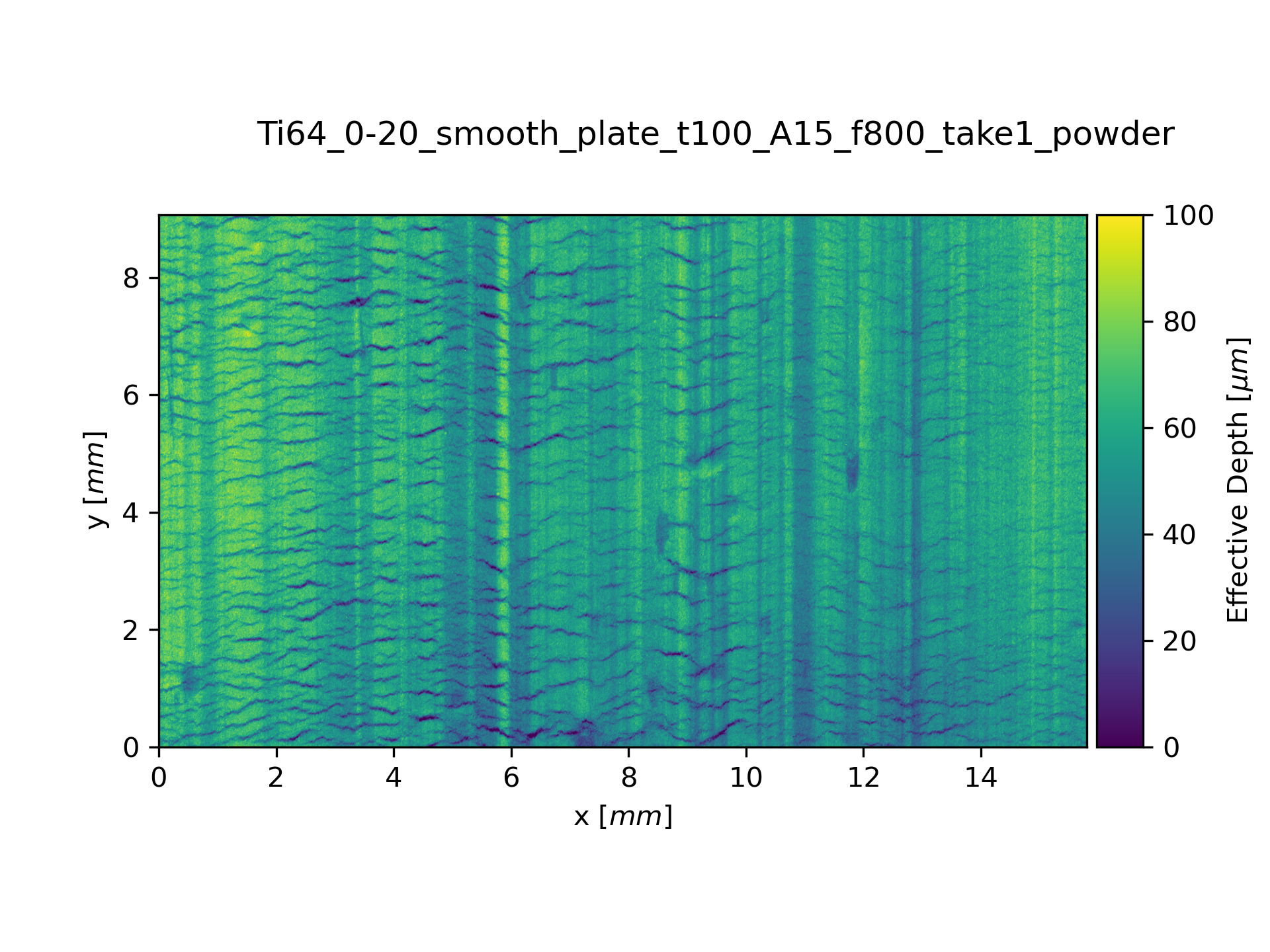}
         \caption{A=15~$\mu m$, f=800~Hz}
         \label{fig:t100_A15_f800}
     \end{subfigure}
     \vfill
     \begin{subfigure}[b]{0.49\textwidth}
         \centering
         \includegraphics[trim = {10mm 15mm 2mm 26mm}, clip, scale=1, keepaspectratio=true, width=\textwidth]{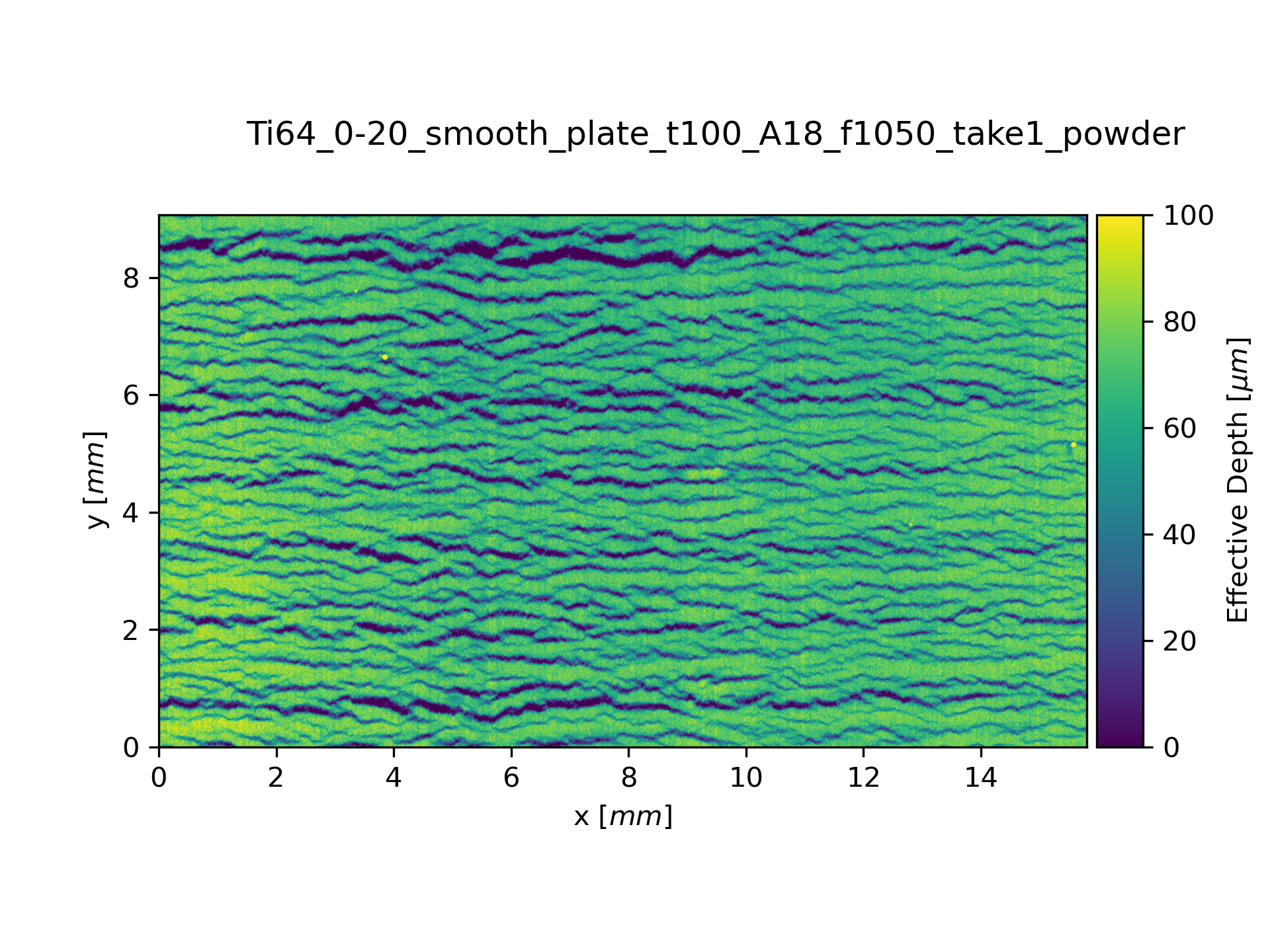}
         \caption{A=18~$\mu m$, f=1050~Hz}
         \label{fig:t100_A18_f1050}
     \end{subfigure}
     \hfill
     \begin{subfigure}[b]{0.49\textwidth}
         \centering
         \includegraphics[trim = {10mm 15mm 2mm 26mm}, clip, scale=1, keepaspectratio=true, width=\textwidth]{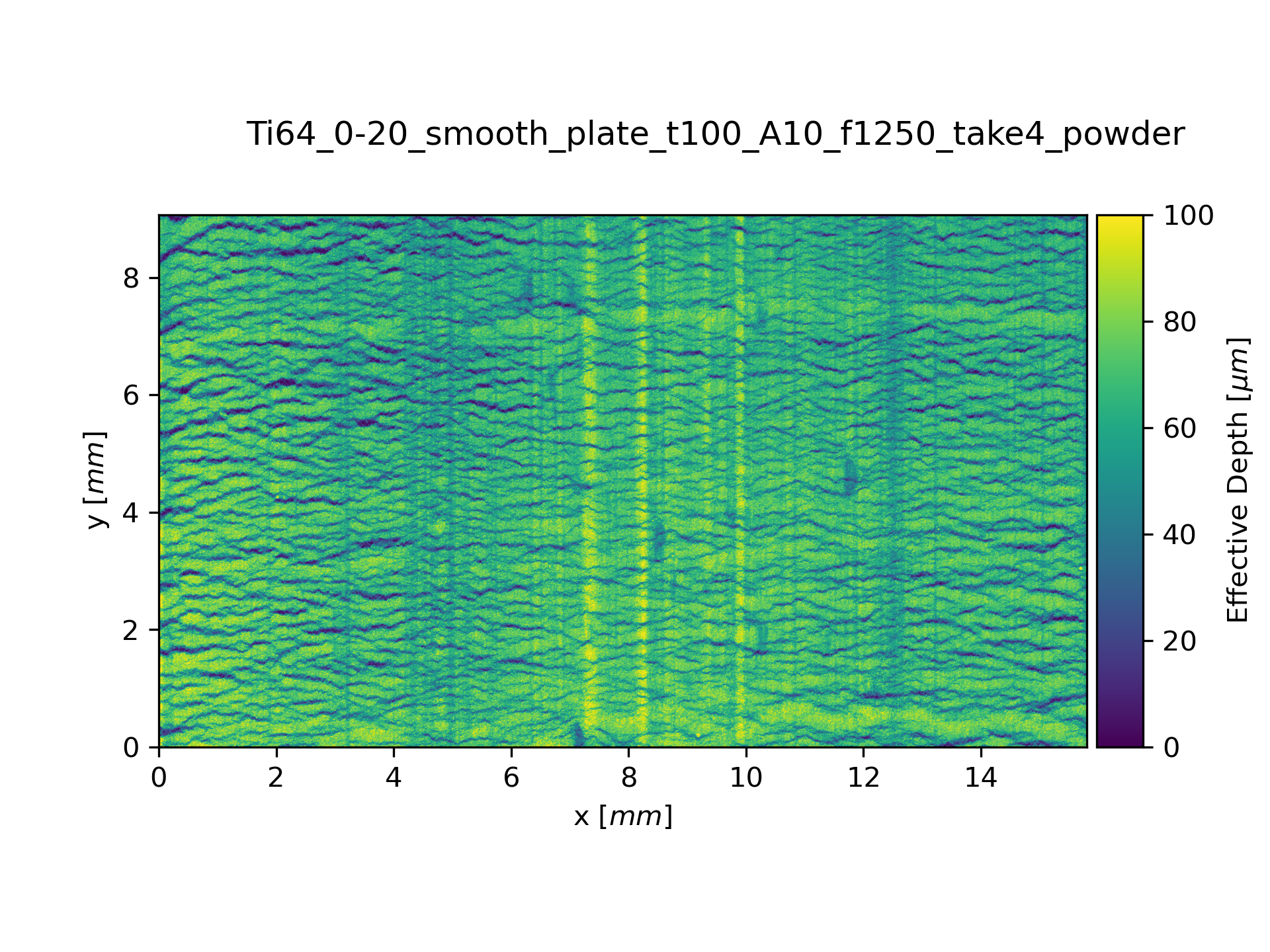}
         \caption{A=10~$\mu m$, f=1250~Hz}
         \label{fig:t100_A10_f1250}
     \end{subfigure}
     \caption{Exemplary X-ray transmission images of powder layers spread via transverse oscillation with a traverse velocity of v=25~$mm/s$ and nominal layer thickness of $t=110~\mu m$; Transmission data converted to effective powder layer depth via radiation transport model as described in~\cite{Penny2021}}
     \label{fig:transverse_exp_100}
\end{figure}

\begin{figure}
     \centering
     \begin{subfigure}[b]{0.49\textwidth}
         \centering
         \includegraphics[trim = {0mm 0mm 0mm 0mm}, clip, scale=1, keepaspectratio=true, width=\textwidth]{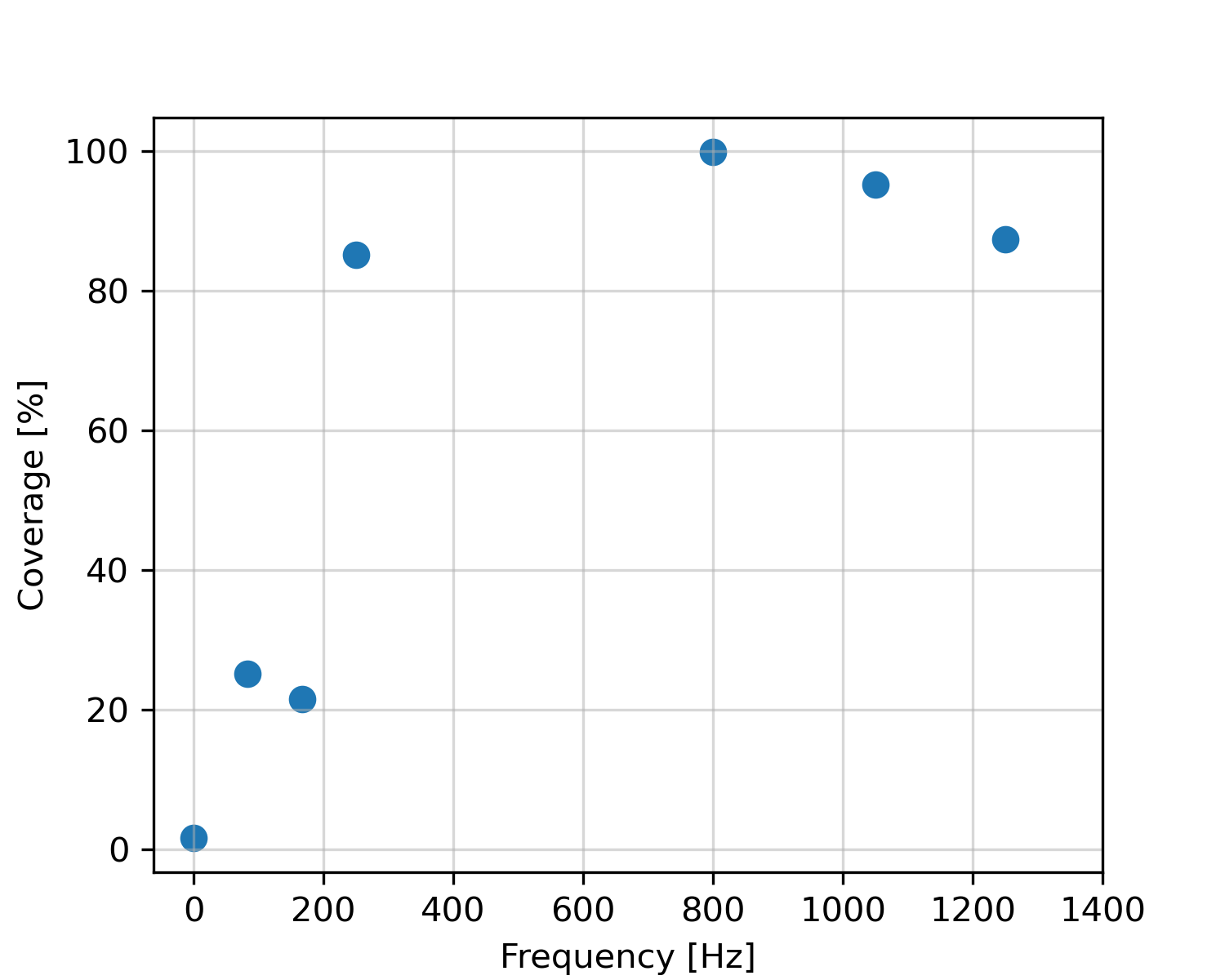}
         \caption{}
         \label{fig:transverse_coverage}
     \end{subfigure}
     \hfill
     \begin{subfigure}[b]{0.49\textwidth}
         \centering
         \includegraphics[trim = {0mm 0mm 0mm 0mm}, clip, scale=1, keepaspectratio=true, width=\textwidth]{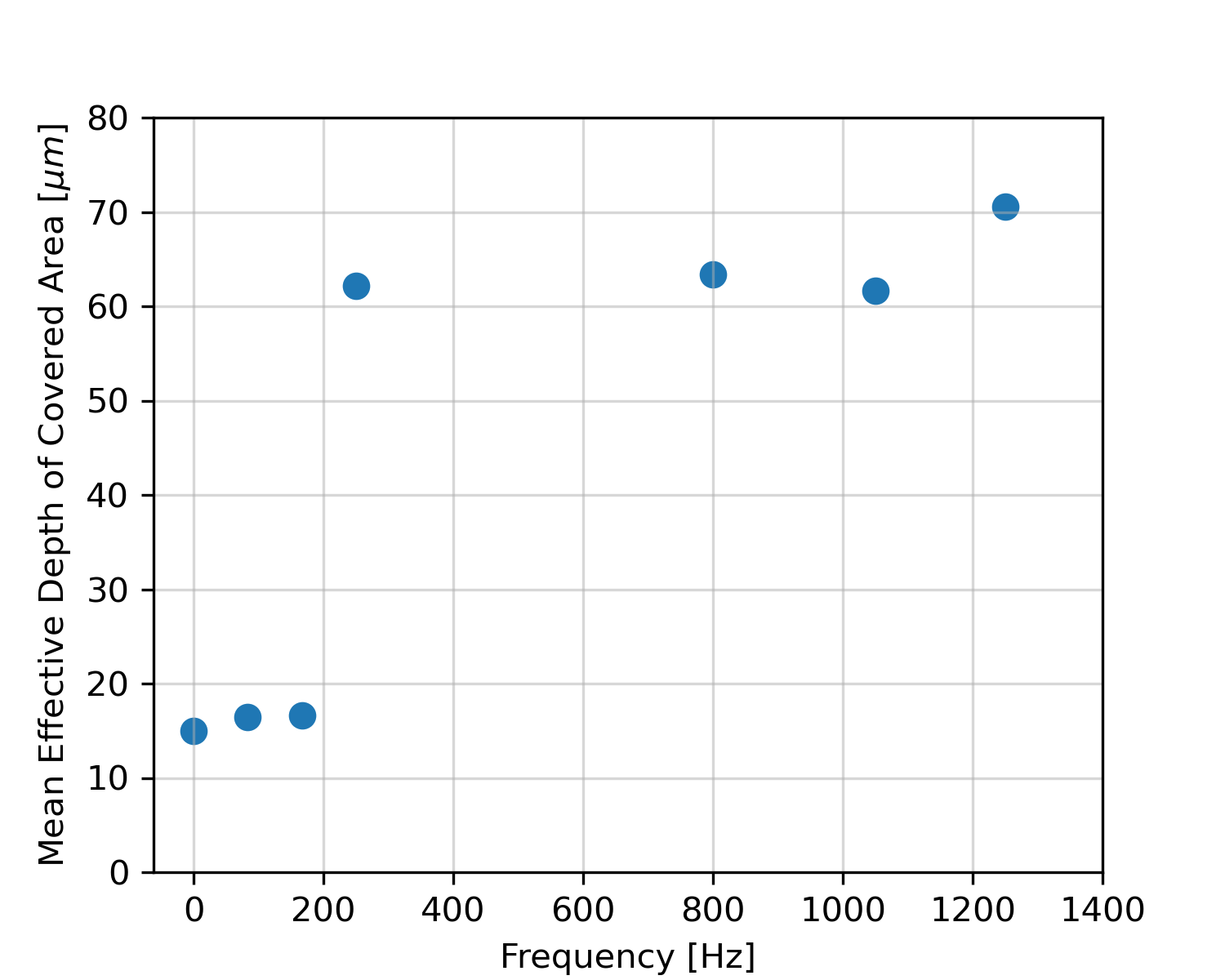}
         \caption{}
         \label{fig:transverse_mean_coverage}
     \end{subfigure}
     \caption{(a) Mean coverage of powder layers, defined as area with an effective depth above $10~\mu m$, and (b) mean effective depth of covered area}
     \label{fig:transverse_coverage_analysis}
\end{figure}

Spreading experiments with transverse oscillation of the non-rotating roller were performed for nominal layer thicknesses of $110~\mu m$ and $300~\mu m$, with roller forward velocity of 25~$\frac{mm}{s}$.
For thickness of 110~$\mu m$, exemplary images of the effective powder layer depth as measured via transmission X-ray imaging are shown in Figure~\ref{fig:transverse_exp_100}, for a range of frequencies from 0~Hz to 1250~Hz. With increasing frequency and therefore equivalent rotational velocity, the effective thickness of the powder layers increases. Without transverse oscillation, essentially no powder is deposited as shown in Figure~\ref{fig:t100_A0_f0}. With increasing frequency, more powder is deposited, and both the density and coverage of the layers visibly increases. Figure~\ref{fig:transverse_coverage} shows the average coverage of imaged powder layers over the range of frequencies. Coverage here is defined as the relative share of the area that has a packing fraction of higher than $10~\mu m$.
However, all layers with a layer thickness of $110~\mu m$ have cracks generally parallel to the roller edge and perpendicular to the spreading direction, even for the highest frequencies. These cracks, which were not observed in the simulations, might be a consequence of unwanted vibrations due to a potentially lack of stiffness of the apparatus as discussed below.

\begin{figure}
     \centering
     \begin{subfigure}[b]{0.49\textwidth}
         \centering
         \includegraphics[trim = {3mm 0mm 8mm 10mm}, clip, scale=1, keepaspectratio=true, width=\textwidth]{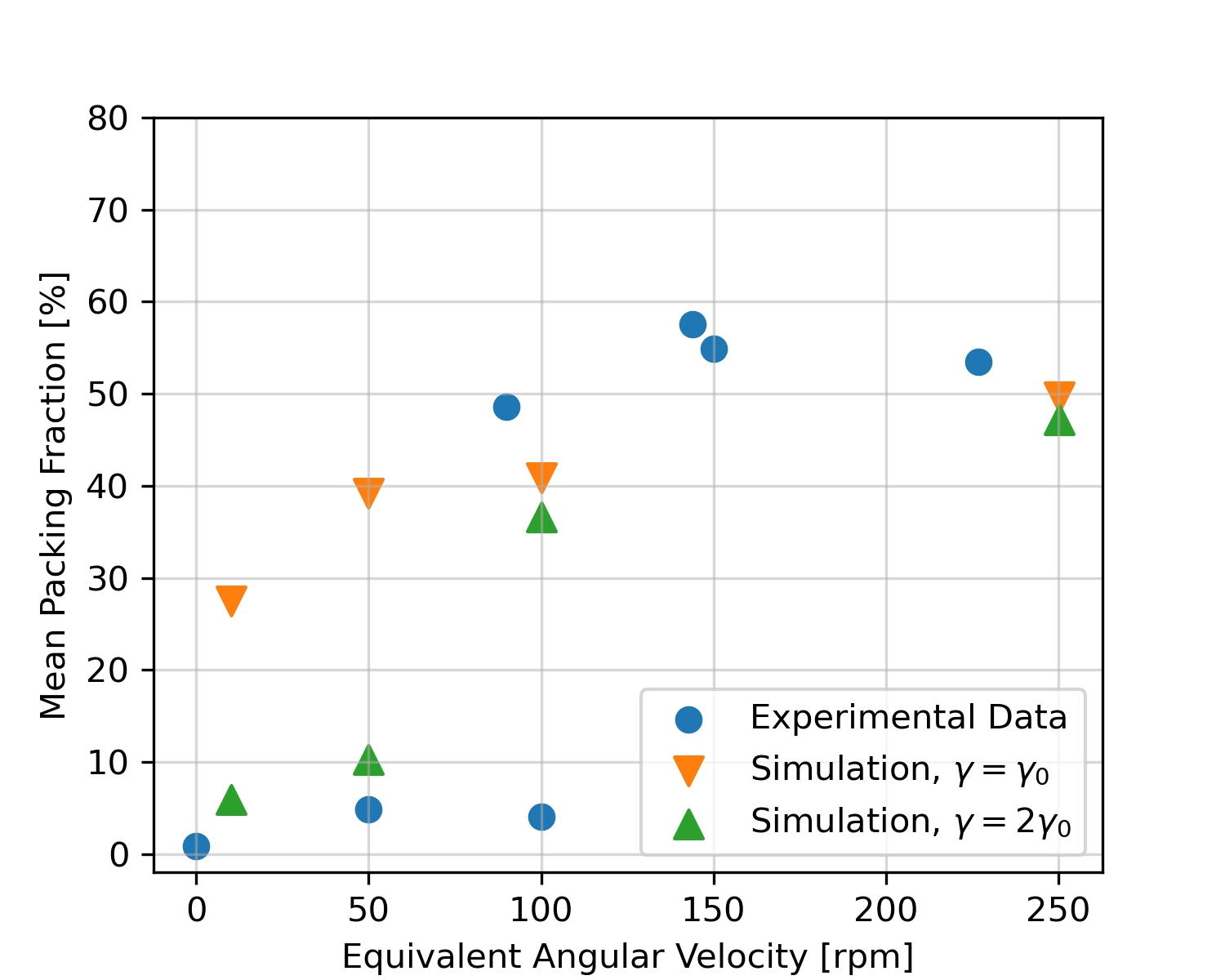}
         \caption{}
         \label{fig:Transverse_eff_depth_rpm_100}
     \end{subfigure}
     \hfill
     \begin{subfigure}[b]{0.49\textwidth}
         \centering
         \includegraphics[trim = {3mm 0mm 8mm 10mm}, clip, scale=1, keepaspectratio=true, width=\textwidth]{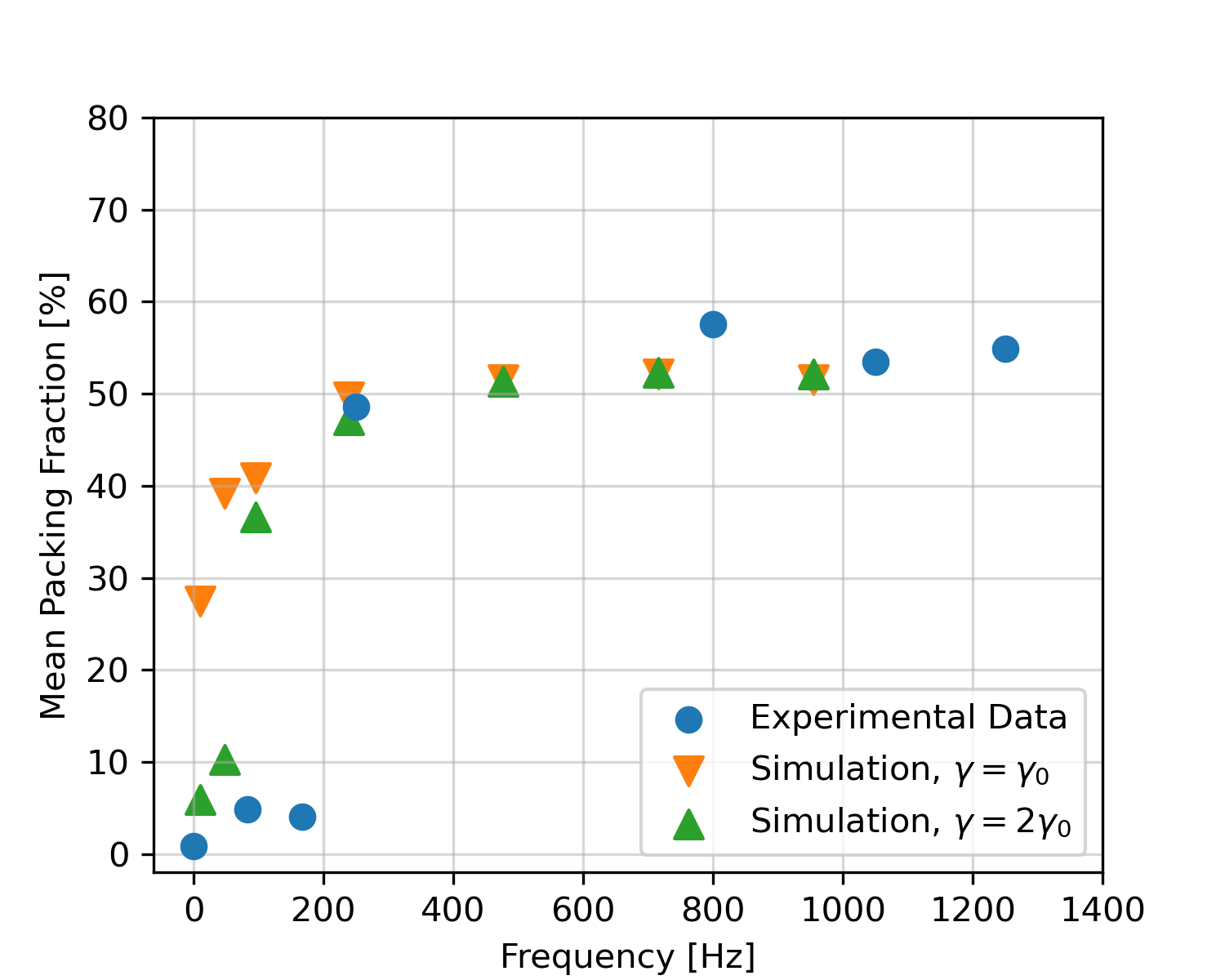}
         \caption{}
         \label{fig:Transverse_eff_depth_f_100}
     \end{subfigure}
     \caption{Comparison of experimental data with simulations, with x-axis represented as (a) equivalent angular velocities, and (b) oscillation frequencies: Mean packing fraction of powder layers, spread with transverse oscillation, for oscillation amplitudes in the range of 0 to 50~$\mu m$ for experimental data points and 87~$\mu m$ for simulated data points}
     \label{fig:Transverse_exp_vs_sim}
\end{figure}

The experimental results for the nominal layer thickness of 110~$\mu m$ are compared to the computational results in Figure~\ref{fig:Transverse_exp_vs_sim}, with mean layer packing fraction plotted against equivalent angular velocity, as well as oscillation frequency.
Experimental data is averaged across all data points recorded for each parameter combination, with the amounts of experiments conducted for the same parameter ranging from 1 to 5 for a total of 20 experiments displayed in the Figures. A lower amount of repeated experiments was conducted for parameter combinations that were challenging to achieve.
For low frequencies / equivalent angular velocities, the powder layer is of very low mean packing fractions below 10~\%. Around 200~Hz frequency, there is a sudden jump in the packing fraction, and all layers spread with a higher frequency have a mean packing fraction of around 50~\% and above. Above that frequency threshold, the powder layer density is relatively consistent, albeit less consistent than the simulations that benefit from consistent and virtually perfectly stable boundary conditions. Notably, experimental layers also achieve very high average densities of around 60~\%. The computational results for $\gamma = \gamma_{fine}$, $\gamma_{substrate}=\gamma$ (compare Figure~\ref{fig:Fig13_a}) predict higher levels of packing fraction for lower frequencies, without the observed sudden jump. When comparing to a more cohesive powder with $\gamma = 2~\gamma_{fine}$, $\gamma_{substrate}=\gamma$ (Figure~\ref{fig:Paper2_comparison_high_adh_norm_padh}), the computational predictions align more closely with the experimental results, with a jump in packing fraction around 150~Hz.
More generally, it has to be noted that a perfect quantitative comparison between simulation and experiment is not to be expected in this work, for reasons such as (i) slightly different powder size distribution (PSD) of the experimental powder, (ii) no explicit modeling of the fine PSD, but implicit modeling using the self-similarity approach, and (iii) uncertainty about experimental factors not considered in the model, such as environmental humidity, stiffness of the equipment, non-idealized surface interaction between powder and surfaces.

\begin{figure}
     \centering
     \begin{subfigure}[b]{0.49\textwidth}
         \centering
         \includegraphics[trim = {10mm 15mm 2mm 26mm}, clip, scale=1, keepaspectratio=true, width=\textwidth]{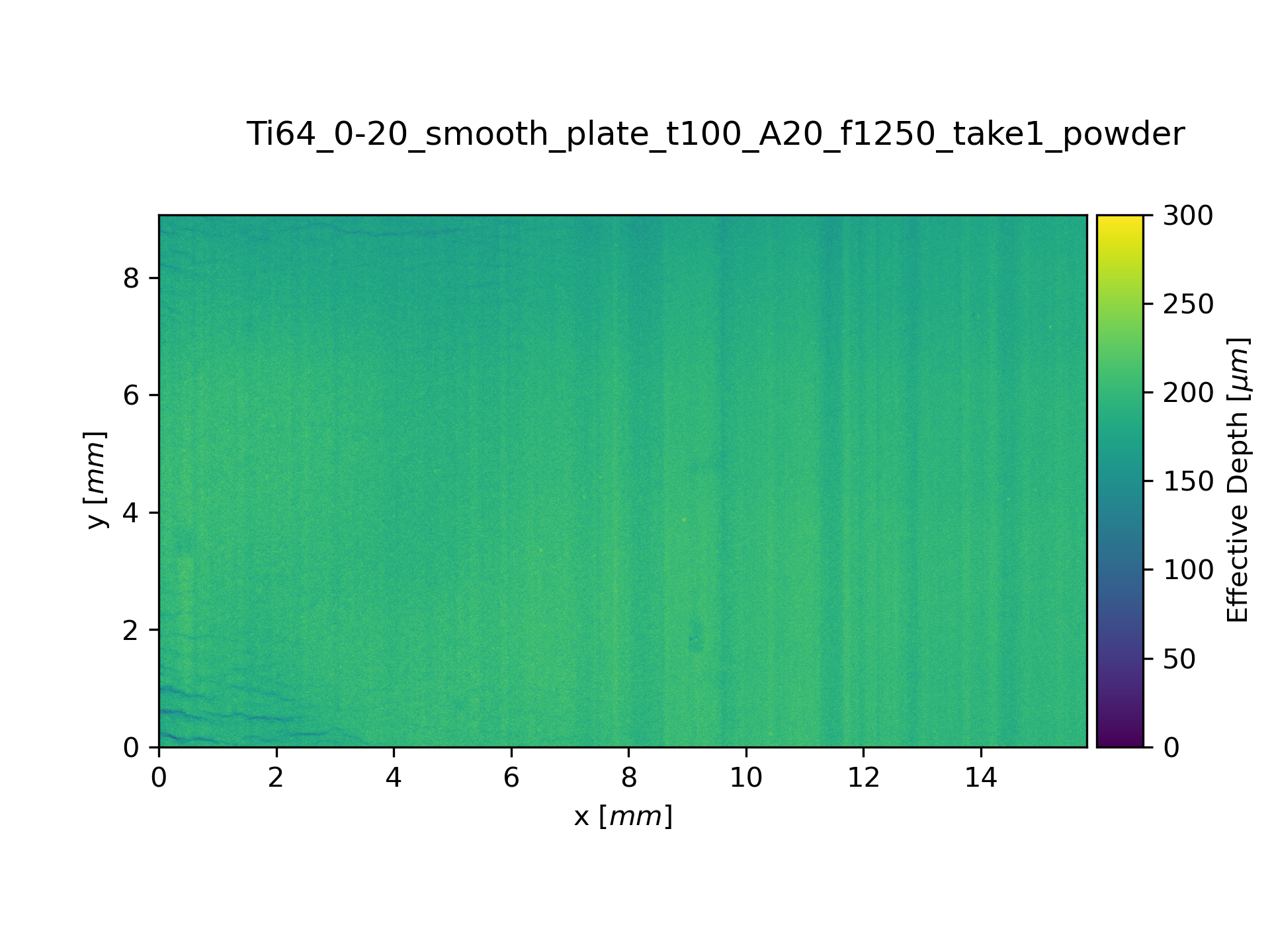}
         \caption{A=20~$\mu m$, f=1250~Hz, Sample 1}
         \label{fig:t300_A20_f1250_1}
     \end{subfigure}
     \hfill
     \begin{subfigure}[b]{0.49\textwidth}
         \centering
         \includegraphics[trim = {10mm 15mm 2mm 26mm}, clip, scale=1, keepaspectratio=true, width=\textwidth]{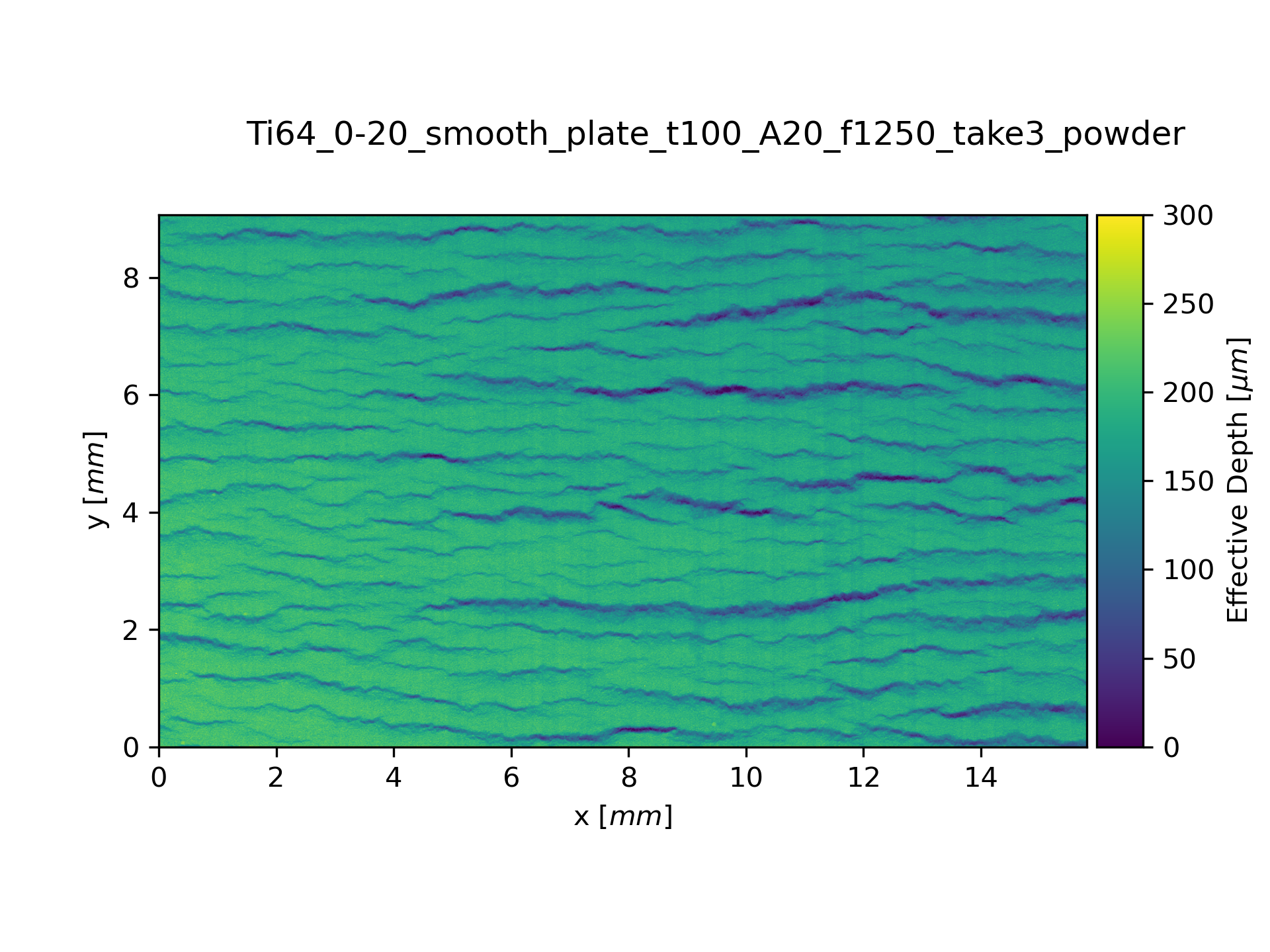}
         \caption{A=20~$\mu m$, f=1250~Hz, Sample 2}
         \label{fig:t300_A20_f1250_3}
     \end{subfigure}
     \vfill
     \begin{subfigure}[b]{0.49\textwidth}
         \centering
         \includegraphics[trim = {10mm 15mm 2mm 26mm}, clip, scale=1, keepaspectratio=true, width=\textwidth]{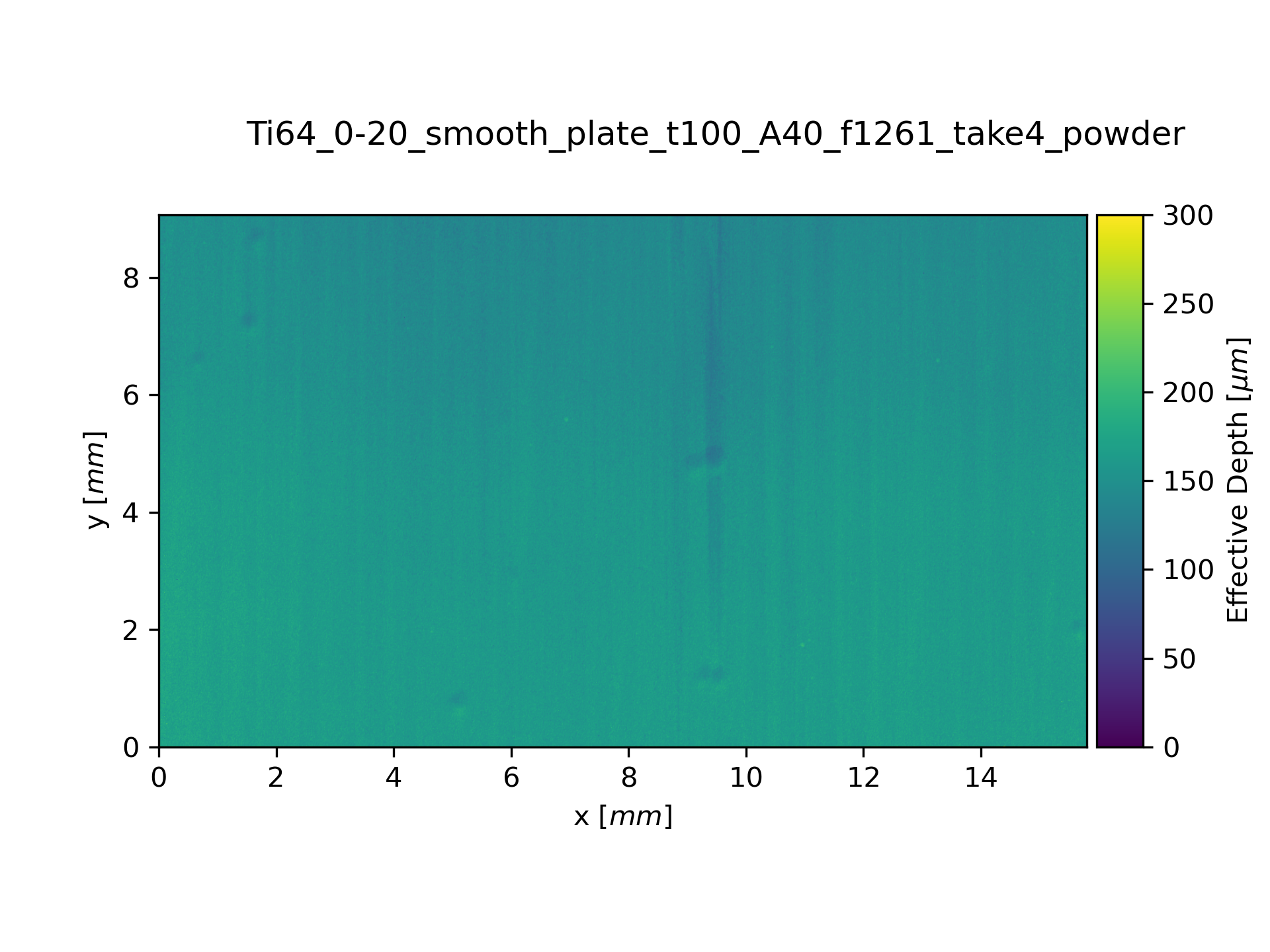}
         \caption{A=40~$\mu m$, f=1261~Hz, Sample 1}
         \label{fig:t300_A40_f1261_4}
     \end{subfigure}
     \hfill
     \begin{subfigure}[b]{0.49\textwidth}
         \centering
         \includegraphics[trim = {10mm 15mm 2mm 26mm}, clip, scale=1, keepaspectratio=true, width=\textwidth]{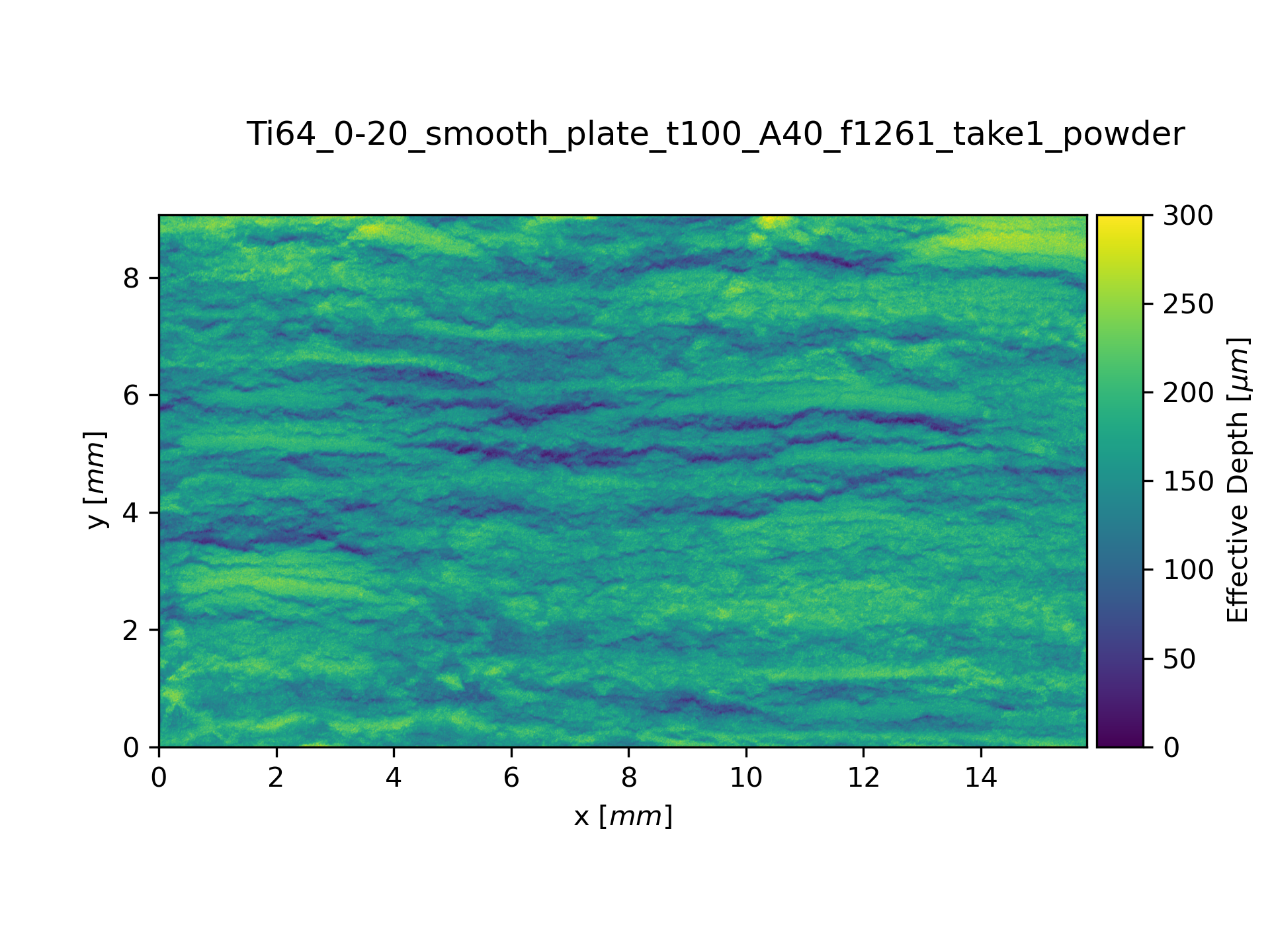}
         \caption{A=40~$\mu m$, f=1261~Hz, Sample 2}
         \label{fig:t300_A40_f1261_1}
     \end{subfigure}
     \caption{Exemplary X-ray transmission images of powder layers spread via transverse oscillation with a traverse velocity of v=25~$mm/s$ and nominal layer thickness of $t=300~\mu m$}
     \label{fig:transverse_exp_300}
\end{figure}

As described earlier, the build piston and build platform are replaced by a thin aluminum build plate in order to allow for X-ray imaging directly on the spreading testbed. However, under certain conditions such as when spreading continuously over the raised edge of the build plate without stopping, oscillations were induced in this structure, causing the thin build plate to vibrate vertically, rendering the spread powder layer unusable. Powder layers where vertical vibration was observed were not included in the data shown here.

This effect is mitigated with a larger gap between the roller and build plate, and therefore a thicker powder layer. Therefore, layers with nominal thickness of 300~$\mu m$ were spread at the highest frequencies, with exemplary results shown in Figure~\ref{fig:transverse_exp_300}. Here, very consistent and uniform powder layers were spread, that only exhibit some streaking behavior, likely caused by surface roughness on the roller. Over the course of the experimental campaign, the nominal layer thickness was repeatedly validated and adjusted with steel feeler gauges that ultimately scratched the surface of the aluminum roller as well as the aluminum build plate, with scratches oriented along the spreading direction that could be responsible for the observed streaking.
While the results shown in Figure~\ref{fig:t300_A20_f1250_1} and Figure~\ref{fig:t300_A40_f1261_4} are very promising, the testbed was not able to consistently achieve this high level of layer quality, likely due to deficiencies in stiffness of the testbed. Figure~\ref{fig:t300_A20_f1250_3} is an example where layer breaking was observed under the same kinematic spreading parameters, and Figure~\ref{fig:t300_A40_f1261_1} is an example of a X-ray transmission image of a layer that experienced vertical vibrations of the build plate and/or roller. 

\begin{figure}
 \begin{center}
   \includegraphics[scale=1, keepaspectratio=true, width=0.75\textwidth]{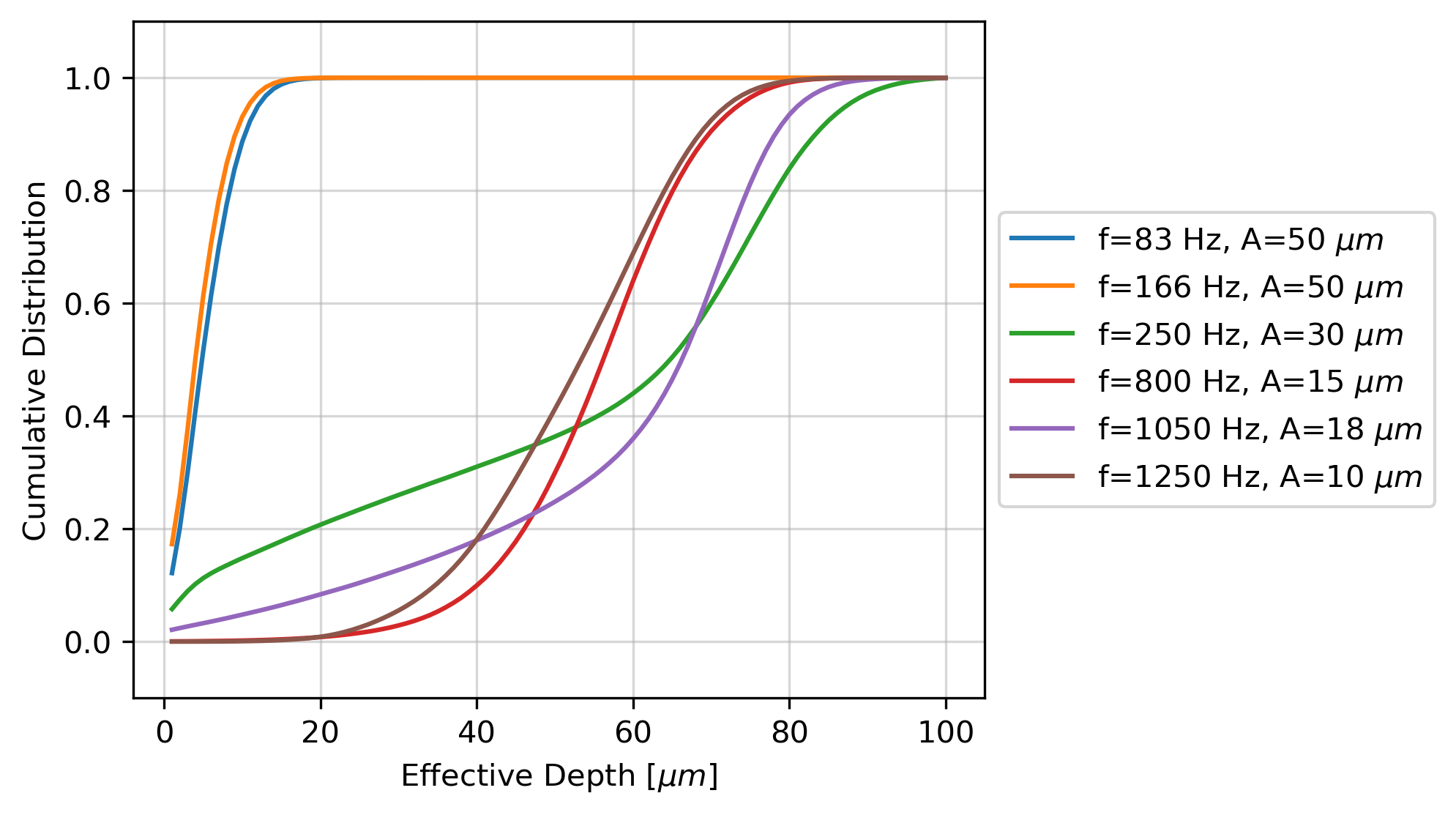}
 \end{center}
 \caption{Cumulative distribution functions of effective depth for exemplary spreading parameters and nominal layer thickness of $110~\mu m$, averaged over samples spread with the same parameters}
 \label{fig:Cum_dist_layers}
\end{figure}

For a more statistical analysis of the layer effective depth, the cumulative distribution function (CDF) of effective layer depth per pixel is analyzed and shown in Figure~\ref{fig:Cum_dist_layers}. Each curve is constructed by evaluating the cumulative sum over a histogram of the effective depth of X-ray transmission images shown above. Images for the same kinematic spreading parameters are averaged. 
The curves for the low frequency regime (83~Hz and 166~Hz) increase sharply at low effective depth. Both curves reach 100~\% before the effective depth of 20~$\mu m$, which means that no pixels have an effective depth above that value. 

The data for 250~Hz is an example of significant layer breaking, which causes the CDF to look similar to a superposition of two Gaussian distributions -- one for the low packing fractions of the cracks in the layer, and one for the areas where powder was deposited with extremely high density. The fact that there are areas with over 80~$\mu m$ effective depth indicates, that either the roller had a slight vertical vibration, or that the nominal layer thickness deviated from the target of 110~$\mu m$. 
Both curves for 800~Hz and 1250~Hz have strong similarities with a normal distribution, although the layer cracking visible in the X-ray images leads to range of pixels with a low effective depth. A narrow normal distribution with a high lower bound would represent an optimal layer with a high level of uniformity and average effective depth.
The shape of the data for 1050~Hz resembles a mixture of the curves for 250~Hz, and 1250~Hz. The similarity with the 250~Hz curve might stem from the similarly large cracks, while the covered areas feature the high local effective depth that is visible in the results for 1250~Hz (compare Figure~\ref{fig:t100_A18_f1050}).

\begin{figure}
 \begin{center}
   \includegraphics[scale=1, keepaspectratio=true, width=\textwidth]{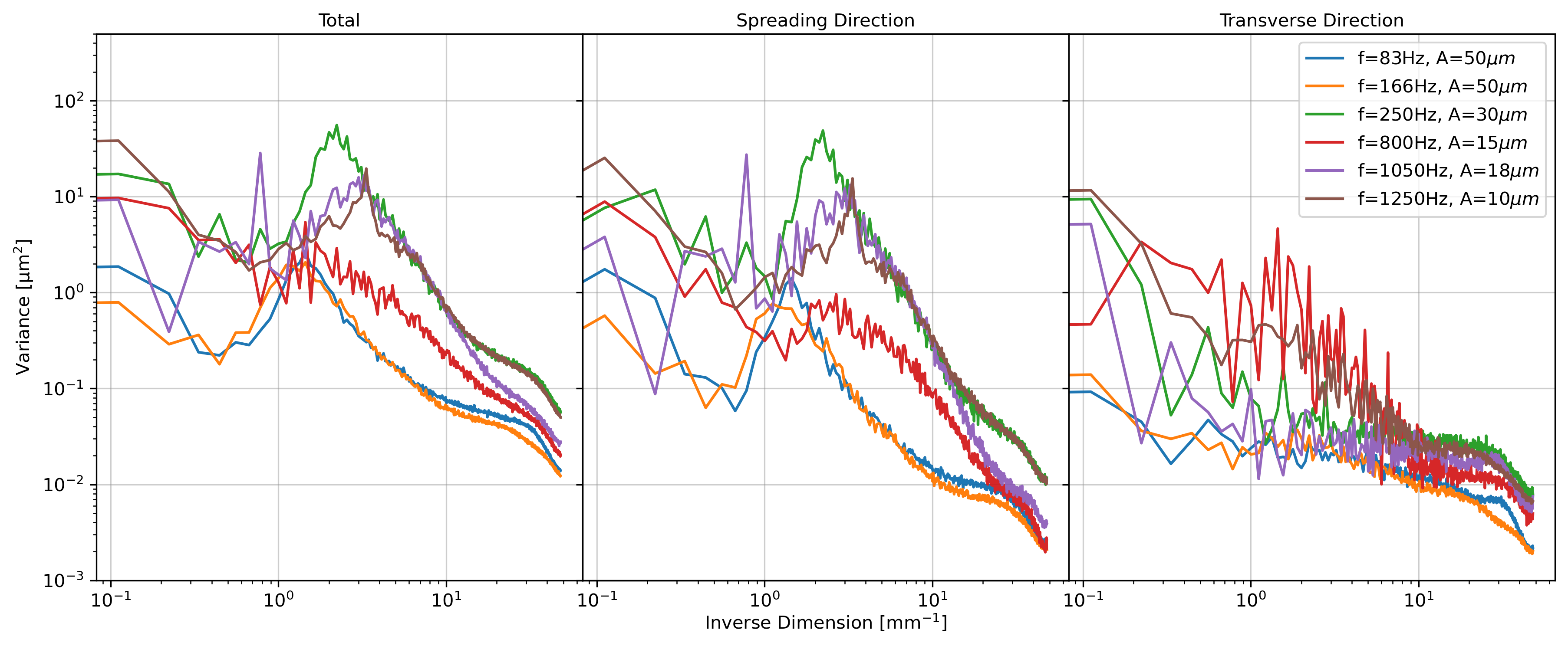}
 \end{center}
 \caption{Analysis of the power spectral density of the effective depth variance of exemplary powder layers spread with transverse oscillation, as measured by X-ray transmission imaging}
 \label{fig:PSD_layers}
\end{figure}

Two types of powder layer defects are visible in the X-ray transmission images: (i) streaking in the spreading direction, and (ii) layer breaking in the transverse direction.
Penny et al. introduced analysis of spatial power spectral density of effective depth maps in \cite{Penny2021}. Power spectral density analysis applies the two-dimensional Fourier transform in the spatial domain instead of the time domain. In doing so, the spatial dimensions get converted into spatial frequencies (or inverse dimensions), in units of inverse length (1/mm). The complex amplitude data returned by the Fourier transform is squared to yield the power spectral density (PSD) as function of inverse dimension (i.e., variance of the powder layer vs. inverse spatial dimension). Finally the variance function is normalized such that integrating over the entire PSD results in the statistical variance of the entire powder layer that is analyzed. 

Within a resulting PSD distribution, the inverse dimension correlates to a radius around the origin of the PSD plot, while also preserving direction. Integrating the PSD distribution in polar coordinates results in all defects mapped to their respective inverse dimensions.
Because direction is preserved, integrating over a section of the circle that is relevant for the direction of interest -- in this case spreading direction and transverse direction -- allows analysis of the defect patterns in these directions. A range of $\pm15^{\circ}$ around these two directions is used for the integration. 

Figure~\ref{fig:PSD_layers} plots PSD analysis of the variance of effective depth versus inverse dimension, of each layer in total (left), in the spreading direction (middle), and transverse to the spreading direction (right).
A data point around $10^{-1}~mm^{-1}$ inverse dimension corresponds to a variance with a period of 10~mm. Since the analyzed image is smaller than 10~mm in side length, that data corresponds to very high-level variations. Conversely, data around $10^{0}~mm^{-1}$ corresponds to a variance with a period of 1~mm, and data around $10^{1}~mm^{-1}$ corresponds to a variance with a period of 0.1~mm.

With help of the total variance plot, the similarities and differences between powder layers spread under different conditions can be assessed. For example, the (averaged) curves for the powder layers spread with 83~Hz, and 166~Hz are very close to each other, as these layers look very similar. The low range of effective depths in these images that was described earlier also results in relatively low absolute variances, which is not necessarily a sign of quality.

Decomposing the variance by direction provides valuable insight into the structure of the different layers. In the transverse direction (right side of Figure~\ref{fig:PSD_layers}), the 800~Hz as well as the 1250~Hz layers exhibit the highest variance in the inverse dimension range of $1-10~mm^{-1}$, corresponding to the significant streaking that is easily detected by bare eye. This is likely not caused by the spreading kinematics, but by wear on the roller surface over the course of the experimental campaign. These defects could potentially be avoided by using a roller made of a material with higher hardness, or a compliant material that is not as prone to scratching.

\begin{figure}
 \begin{center}
   \includegraphics[scale=1, keepaspectratio=true, width=0.5\textwidth]{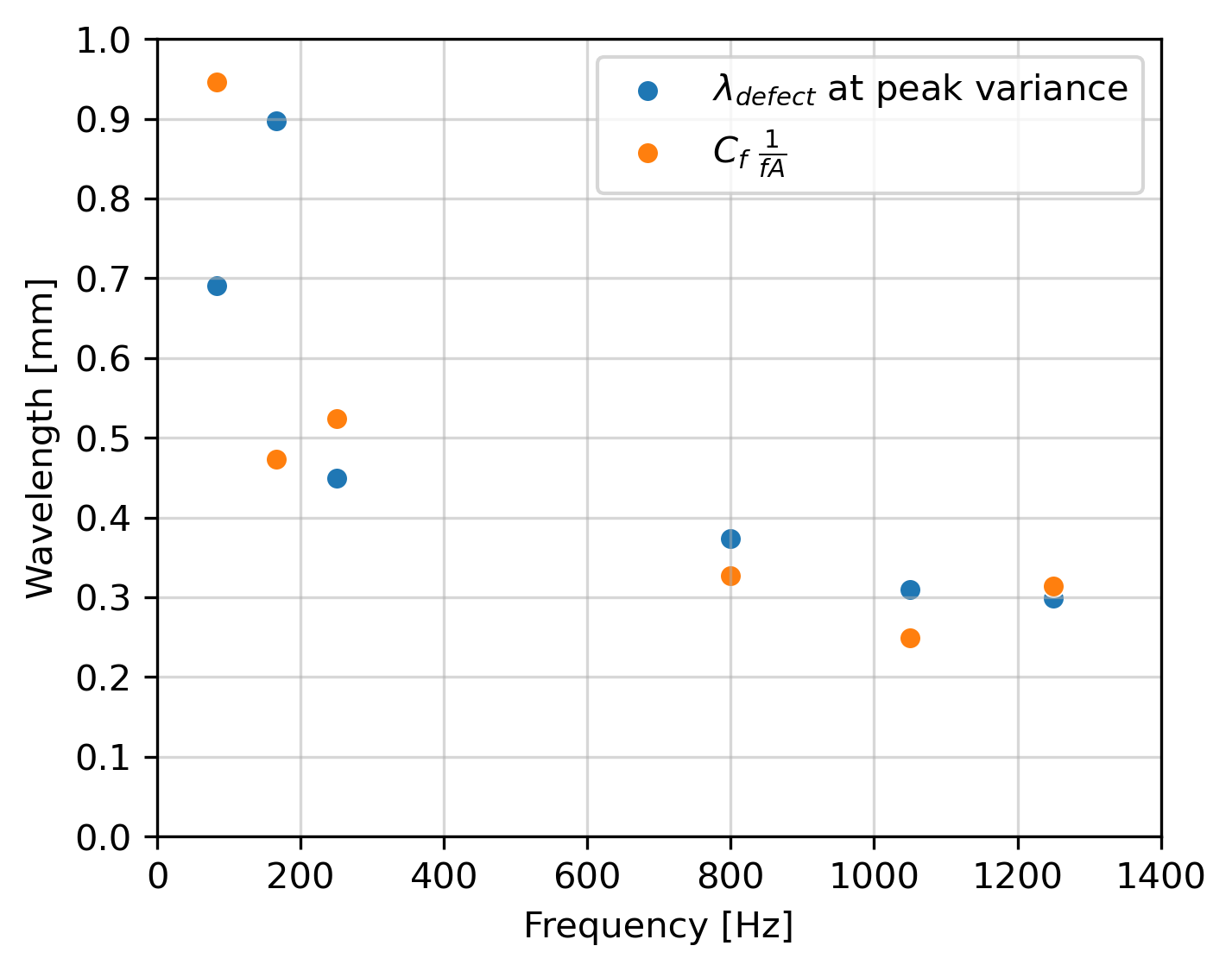}
 \end{center}
 \caption{Calculated wavelength of maximal effective depth variance based on power spectral density analysis of spread powder layers as imaged with X-ray transmission imaging, compared to a theoretical scaling relationship}
 \label{fig:defect_wavelength_freq}
\end{figure}

The analysis in the spreading direction (middle of Figure~\ref{fig:PSD_layers}) corresponds to the breaking of the powder layer and is expected to be caused by the transverse oscillation kinematic. Again the focus is on the range from $1-10~mm^{-1}$ as the key area of interest and a first finding is that the majority of the total variance that is shown in the left plot is stemming from the layer breaking in the spreading direction. The data for each layer has a peak in that range, before consistently tapering off. These peaks of the variance are expected to be the dominant spatial frequency of layer breaking and potentially influenced by the transverse oscillation.

To understand how the spatial character of the layers relates to the oscillation parameters,  wavelengths $\lambda_{defect}$ at peak variance are plotted against their corresponding transverse oscillation frequencies in Figure~\ref{fig:defect_wavelength_freq}. There is a clear trend of a decreasing wavelength as the transverse oscillation frequency increases.
Each of the data points for oscillation frequency corresponds to a varying amplitude as described earlier. When considering the corresponding amplitudes in combination with the frequencies, the data points closely follow a fit of $C_f~\frac{1}{fA}$.
For the data at hand, with traverse spreading velocity of $v=25~\frac{mm}{s}$, the identified frequency constant $C_f$ is equal to $C_f = 1e3 ~2\pi v^2$. Since only one spreading velocity $v$ was studied, this scaling is speculative in nature and might be a coincidence -- verification of the constant could be the focus of future work.
Considering that a low wavelength (high frequency) indicates a higher quality of the powder layer, understanding the impact of the spreading parameters is valuable.
For the counter rotation and angular oscillation, it was shown that surface velocity is the critical factor for powder layer quality.
Surface velocity of the roller for the transverse oscillation is determined by $fA$ with oscillation frequency $f$ and amplitude $A$. An increase in either thus results in an improvement of the powder layer quality. 
Further, following the potential fit identified for the constant $C_f$, it appears that an increase in traverse spreading velocity $v$ might result in an increase in the wavelength. As shown previously, an increase in spreading velocity for counter-rotating rollers corresponds in a decrease in powder layer density. It is thus reasonable that an increase in traverse velocity $v$ would also lead to a decrease in powder layer quality for the transverse oscillation kinematic.

Reflecting on the results, a variety of limitations of the experimental testbed were identified. First, the required pre-loading of the springs that allow oscillatory motion of the roller close to resonance frequency was done using only one stiffening bar and the brackets that attach the roller to the lead screws (see Figure~\ref{fig:Oscillation mechanism}). Application of sufficient pre-load and alignment of the roller with the brackets and linear bearings was challenging. Misalignment and locking in the bearings was likely a central reason preventing high oscillation amplitudes and higher frequencies (which require higher pre-loading forces with stiffer springs).
Next, unwanted vertical vibrations of the roller and the build plate were a challenge that at times prevented the formation of uniform layers. The roller would start vibrating when moving over the step up onto the raised build plate, also in part due to the alignment and stiffness limits mentioned earlier.
Last, it is critical to level the carriage relative to the build platform and ensure that the nominal layer thickness is correct across the entire area, which is performed via micrometer screws on the corners of the lead screws. This process is highly iterative and tedious and requires frequent validation via feeler gauges which over time was visible as wear on the roller. 

Furthermore, it is notable that the computational results are not designed for precise quantitative predictions, but primarily to observe qualitative behavior (informed by quantitative results) and understand the underlying physics. The primary reason for that is the employed self-similarity approach for modeling particles and powder cohesion. Due to that approach, the computational domain features larger particles relative to the dimensions of the powder bed and spreading implement as compared to the experimental setup. Among others, this is likely the cause that layer breaking is not observed during the computational results -- in the experiment, a higher number of particles can form particle clusters within a certain fixed volume as compared to the simulation.

Moreover, prior findings~\cite{PENNY2022Blade} suggest that employing a rubber-coated tool could serve as a viable option for preventing streaking defects. Recent simulation outcomes indicate that the oftentimes used O-ring stock possesses enough stiffness to effectively respond to the high-frequency oscillations necessary for distributing the highly cohesive powder~\cite{weissbach2024exploration}.

The fact that layer breaking was mitigated using thicker layers is also an indicator that during a multi-layer spreading experiment, breaking might not be as prevalent as observed during the single-layer experiments conducted in this study. During a real LPBF experiment, the majority of the powder layer is spread on top of powder as compared to a smooth metal surface which might be beneficial. 
Many commercial machines apply hoppers to disperse the powder before spreading. This has been shown to be beneficial \cite{PENNY2022Roller} for spreading very cohesive powders, but no significant difference was observed during experiments conducted by the authors, but has not been explored in full detail. As second potentially promising avenue for future work, sand blasting of the build plate might improve spreading results due to the increased unstructured surface roughness.

\section{Conclusion}

This work describes the computational investigation of a novel powder spreading approach for fine, cohesive powders via transverse oscillation kinematics. Based on the computational results, an experimental powder spreading testbed was modified to accommodate this new actuation, and an experimental parameter study was conducted covering a wide range of frequencies. X-ray transmission images of powder layers were analyzed statistically and by transforming the pixel data into the spatial frequency domain. 
For transverse oscillation frequencies above 200~Hz, powder layers of high packing fraction (between 50-60~\%) were formed. However, thin powder layers feature fine cracks perpendicular to the spreading direction, and statistical analysis revealed that an increase in frequency and amplitude results in finer breaking, presumably until a completely coherent layer is formed.
For an increased nominal layer thickness, continuous and uniform layers of high density were successfully formed, showing the potential of this new spreading approach.


\section{Acknowledgements}

Financial support at MIT was provided by a MathWorks MIT Mechanical Engineering Fellowship (to R.W.). P.P., C.M., and W.W. acknowledge funding by the Deutsche Forschungsgemeinschaft (DFG, German Research Foundation) within project 414180263. C.M. acknowledges the financial support from the European Research Council through the ERC Starting Grant ExcelAM (project number: 101117579).

\newpage
\bibliographystyle{elsarticle-num}
\typeout{}
\bibliography{refs.bib}

\end{document}